\documentclass[aps,prl,amsmath,amssymb,superscriptaddress,twocolumn,showpacs,longbibliography,floatfix,notitlepage]{revtex4-1}

\DeclareSymbolFontAlphabet{\amsmathbb}{AMSb}%
\DeclareSymbolFontAlphabet{\mathbb}{AMSb}
\usepackage[bbgreekl]{mathbbol}

\usepackage{graphicx}
\usepackage{dcolumn}
\usepackage[colorlinks=true,linkcolor=blue ,citecolor=blue,urlcolor=blue]{hyperref}
\usepackage{multirow}
\usepackage[usenames,dvipsnames]{xcolor}
\usepackage{soul}
\usepackage{braket}
\usepackage{bm}
\usepackage{nicefrac}
\usepackage{siunitx}
\usepackage{color}
\usepackage[utf8]{inputenc}   
\usepackage{upgreek}
\usepackage{ulem}
\usepackage{comment}

\usepackage{float}

\newcommand{\VEC}[1]{\boldsymbol{#1}}

\begin{document}

\title{Origin of Incommensurate Magnetic Order in Rare-Earth Magnetic Weyl Semimetals}

\author{Juba Bouaziz}
\email{j.bouaziz@fz-juelich.de}
\affiliation{Peter Gr\"{u}nberg Institut and Institute for Advanced Simulation, 
Forschungszentrum J\"{u}lich \& JARA, D-52425 J\"{u}lich, Germany}
\author{Gustav Bihlmayer}
\affiliation{Peter Gr\"{u}nberg Institut and Institute for Advanced Simulation, 
Forschungszentrum J\"{u}lich \& JARA, D-52425 J\"{u}lich, Germany}
\author{Christopher E. Patrick}
\affiliation{Department of Materials, University of Oxford, Parks Road, Oxford OX1 3PH, United Kingdom}
\author{Julie B. Staunton}
\affiliation{Department of Physics, University of Warwick, Coventry CV4 7AL, United Kingdom}
\author{Stefan Bl\"ugel}
\affiliation{Peter Gr\"{u}nberg Institut and Institute for Advanced Simulation, 
Forschungszentrum J\"{u}lich \& JARA, D-52425 J\"{u}lich, Germany}

\date{\today}

\begin{abstract}
We investigate rare-earth magnetic Weyl semimetals through first-principles 
simulations, analyzing the connection between incommensurate magnetic order and the 
presence of Weyl nodes in the electronic band structure. Focusing on PrAlSi, NdAlSi, 
and SmAlSi, we demonstrate that the reported helical ordering does not originate from 
the nesting of topological features at the Fermi Surface or the Dzyaloshinskii-Moriya 
interaction. Instead, the helical order arises from frustrated isotropic short-range 
superexchange between the $4f$ moments facilitated by $pd$-hybridization with the main
group elements. Employing  a spin Hamiltonian with isotropic exchange and single-ion 
anisotropy we replicate the experimentally observed helical modulation.
\end{abstract}
 
\maketitle

Magnetic Weyl semimetals form an exciting class of topological materials~\cite{kumar2020topological}, 
owing to the possibility they offer of  combining  nontrivial topology in both reciprocal momentum space and the space of magnetic order parameters. This raises the prospect of identifying new topological 
invariants which characterize novel and intriguing physical response 
properties. In momentum space, Weyl nodes emerge as topologically nontrivial points of contact 
or crossings between two Kramers' degeneracy-lifted bands, acting as sinks and sources of diverging Berry curvature~\cite{hasan2021weyl,yan2017topological}. These  correspond to magnetic monopoles strongly affecting 
electronic response and  transport properties, such as  anomalous Hall~\cite{destraz2020magnetism} 
or Nernst effects, magnetoresistance~\cite{huang2015observation,Knoll2020} or optical properties~\cite{nagaosa2020transport}, 
if  they are in the vicinity of the Fermi surface (FS). These emergent Weyl fermions are either a consequence 
of time-reversal symmetry or spatial-inversion symmetry breaking (ISB) in presence of
the spin-orbit interaction (SOI)~\cite{Knoll2022}.

Weyl materials are of great interest magnetically owing to the  complex magnetic textures such as spin-spirals they can exhibit which can be turned into topological nontrivial 
textures such as skyrmions with the application of external magnetic fields. ISB, SOI, and magnetism are also the necessary ingredients  for chiral Dzyaloshinskii-Moriya magnetic
interactions (DMI)~\cite{Dzyaloshinsky:1958,Moriya:1960}. DMI
can compete with exchange interactions, giving rise to chiral spin-spiral ground states~\cite{Dzyaloshinsky:1964} 
and chiral magnetic skyrmions ~\cite{Bogdanov:1989} when magnetic fields are applied. This 
hints at a relation between the Weyl points and skyrmion 
formation. 

All the more surprising is the recent observation of spiral magnetism in 
magnetic rare-earth ($R$) $R$AlSi Weyl semimetals and an apparent link to nesting between topologically 
nontrivial Fermi surface pockets ~\cite{gaudet2021weyl,yao2023large,lyu2020nonsaturating,puphal2020topological,yang2021noncollinear}. 
This suggests that the Weyl points are directly related to the much stronger exchange interaction, 
rather than the weaker DMI, and are the determinants for the magnetic phases. A correlation between band 
structure topology, \textit{i.e.}\ the Weyl points, and the emergence  of an incommensurate magnetic 
order has been found for NdAlSi~\cite{gaudet2021weyl}. The observed helical magnetic order is characterized 
by a wave vector $\boldsymbol{q}$ that matches the vector connecting the topological features observed 
in the FS. The incommensurate order in NdAlSi transforms to a commensurate ferrimagnetic one~\cite{gaudet2021weyl}
at low temperatures, attributed to the magnetic anisotropy originating from crystal field effects~\cite{mackintosh1991rare}. 
For SmAlSi~\cite{yao2023large} helical magnetic order has been identified to persist to lower 
temperatures and include a topological Hall effect characteristic of the $A$-phase~\cite{Neubauer2009} 
in skyrmion materials. PrAlSi exhibits both ferromagnetic and also possibly spin glass or ferromagnetic 
cluster glass behavior~\cite{lyu2020nonsaturating}. Thus, $R$AlSi compounds have garnered significant interest to be 
ideal systems for the exploration of the interplay between nontrivial valence band structures and chiral magnetic 
textures.

\begin{figure*}
 \centering
\includegraphics[width=1\textwidth]{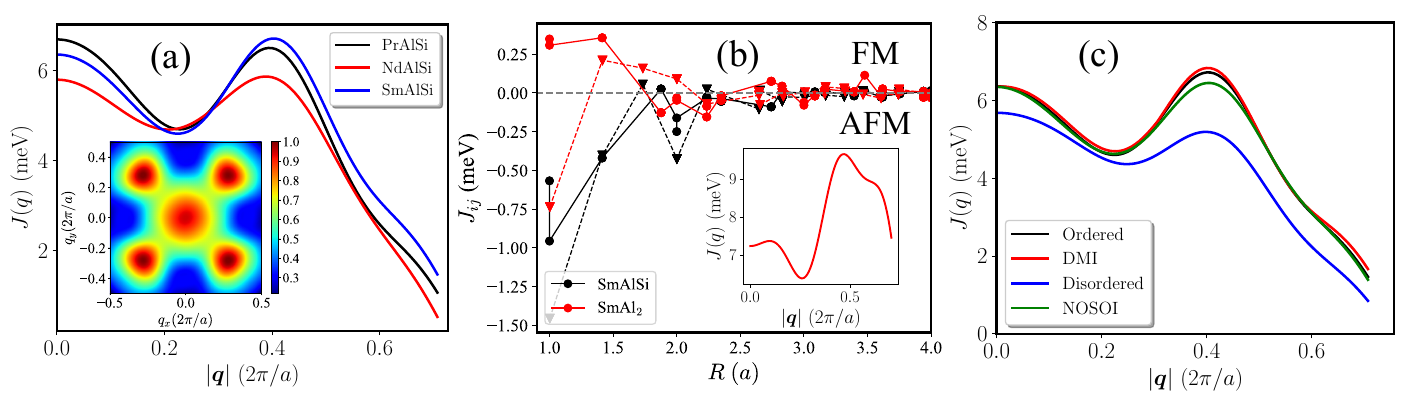}
\vspace{-0.7cm}
\caption{(a) The maximal eigenvalues $J(\boldsymbol{q})$, indicating a competition between ferromagnetic 
($q=0$) and helical order at $q_p$ in $[110]$ direction, for PrAlSi, NdAlSi, and SmAlSi. 
The inset highlights the $4$-fold degenerate peak for SmAlSi. (b) Real-space isotropic exchange interactions 
in SmAlSi and SmAl$_2$ between $4f$ atoms at distance $R$, showing different regimes of SX and RKKY.
The full (dotted) lines represent intra-layer (inter-layer) interactions, the inset depicts the 
$J(\boldsymbol{q})$ for SmAl$_2$. (c) A comparison of $J(\boldsymbol{q})$ for different approximations: 
SmAlSi compound with ordered Si and Al atoms (black), including the DMI (red), without SOI (green), and with a
disordered Si and Al distribution, restoring centrosymmetry,  with SOI (blue).}
\vspace{-0.5cm}
\label{fig_all_REAlSia}
\end{figure*}

Although the relation of FS nesting vectors to helical spin wave vectors is suggestive, hard quantitative evidence which relates the Weyl points to the collective phenomenon of a magnetic phase is missing. This concern is more apposite when one considers the many known exchange interactions between  local magnetic moments in solids, \textit{e.g.}\  superexchange (SX)~\cite{Anderson1950,Solovyev1999}, double exchange~\cite{Solovyev1999_2}, and indirect exchange or Ruderman-Kittel-Kasuya-Yosida (RKKY)  interactions~\cite{hughes2007}.  In particular, long-ranged competing RKKY interactions~\cite{hughes2007,Mendive2017}, which are exactly derived from FS features such as nested sheets~\cite{Bouaziz2022}, can lead to incommensurate magnetic order, and ultimately to complex spin textures.

In this letter,  for  three specific candidates, PrAlSi, NdAlSi, and SmAlSi, 
where the rare-earth compounds are not metals, but semi-metals, we explore to what extent 
features of their threadbare FSs, such as Weyl points, can also be major contributors to the 
cause of the materials' magnetic orders. Moreover, we also assess whether short-ranged but frustrated 
SX interactions are preeminent owing to the presence of a quasi-band gap. To this end we examine the magnetic interactions 
between the rare-earth atoms given by a first-principles account, which is unbiased as to the mechanism. 

In order to extract some generic insights about the magnetic interactions prevalent in these 
$4f$ magnetic Weyl semi-metal materials, we study a GdAlSi prototype (supplementary note 1) whose lattice parameters are set  to match 
each $R$AlSi material~\cite{hughes2007,Mendive2017}. This enables us to discriminate 
between RKKY-like magnetic interaction mechanisms, reliant on electronic structure near the 
Fermi energy in metals, or SX-like mechanisms inherent in magnetic insulators. We 
find generic competing ferromagnetic and incommensurate helical order interactions. We then 
perform {\it ab initio} crystal field theory 
calculations of single-ion anisotropies~\cite{Patrick2019} to determine the specific magnetic 
ordered structures.

We employ Hubbard $U$-corrected density functional theory (DFT+$U$)~\cite{Dudarev2019} 
calculations to investigate the magnetic interactions and electronic band structures. The calculations are performed with the 
all-electron full-potential Korringa-Kohn-Rostoker (KKR) Green function method~\cite{papanikolaou:2002},
including scalar relativistic effects and spin-orbit coupling self-consistently~\cite{Bauer:2014}. 
We compute the magnetic interactions between the $4f$ magnetic atoms using the infinitesimal 
rotation method~\cite{liechtenstein1984,Ebert:2009,Solovyev:2021}. The real space exchange 
interactions tensor and the corresponding lattice Fourier transforms are carefully 
inspected, unveiling the different exchange mechanisms at play. The lattice constants employed 
in the calculations are taken from experiment (supplementary note 2). 
 
A minimal spin Hamiltonian, $\mathcal{H}_\text{m}$, for a non-centro\-symmetric crystal,  with  the magnetic interactions, which we have calculated {\it ab initio}, is given by:
 \begin{equation}
\begin{split}
\mathcal{H}_\text{m}
 = & -\frac{1}{2}\sum_{i\ne j}J_{ij}\,\VEC{m}_{i}\cdot\VEC{m}_{j}
+\frac{1}{2}\sum_{i\ne j}\VEC{D}_{ij}\cdot(\VEC{m}_{i}\times\VEC{m}_{j})\\
& -\sum_{i}K_{i}\,(\VEC{e}_{n}\cdot\VEC{m}_{i})^2\quad.
\end{split}
\label{magn_ham}
\end{equation}
$\VEC{m}_{i}$ is the direction of the magnetic moment at a $R$-site $i$, and $\VEC{e}_{n}$ 
is the direction of the (effective) easy axis specified with respect to the crystal structure. The first 
term refers to the isotropic Heisenberg interactions, the second  the anti-symmetric 
Dzyaloshinskii–Moriya interactions (DMI)~\cite{Dzyaloshinsky:1958,Moriya:1960}, which promote 
chiral spin textures~\cite{heinze2011spontaneous}, and the third the crystal field single-ion 
anisotropy at the rare-earth sites ($K_{i}$). We can safely neglect the two-ion anisotropy 
due to its small size compared to $K_{i}$. In order to find the origin of the helical order 
observed in the $R$AlSi family, we inspect the Fourier transform 
${\mathcal{J}}^{\alpha\beta}_{mn}(\boldsymbol{q})$ of the magnetic exchange interactions.
$m$ and $n$ denote the atomic indices in the unit cell labelling the two magnetic atoms 
in $R$AlSi, $\alpha$ and $\beta$ indicate the $\{x,y,z\}$ components. The maximal eigenvalue $J(\boldsymbol{q})$ of the Fourier transform matrix ${\mathcal{J}}^{\alpha\beta}_{mn}(\boldsymbol{q})$  provides information on the magnetic 
order and an estimate of the transition temperature $T_\text{N}$. $J(\boldsymbol{q})$ is 
depicted in Fig.~\ref{fig_all_REAlSia}(a) for our Gd prototype with the lattice structures of PrAlSi, NdAlSi 
and SmAlSi, respectively (supplemental note 3). In each case, we see two peaks of roughly 
comparable magnitude at $\boldsymbol{q}=0$, indicative of intra-layer ferromagnetic correlations, and at 
$\boldsymbol{q}= \frac{2\pi}{a} (q,q,0)$ with $q \approx 0.3$ ($a$ being the in-plane lattice 
constant), which describes single-$q$ spiroidal magnetic correlations~\cite{Mendive2017,Bouaziz2022}. The dominant peak 
determines the magnetic order that will form below the transition temperature.

\begin{figure*}
 \centering
\includegraphics[width=1\textwidth]{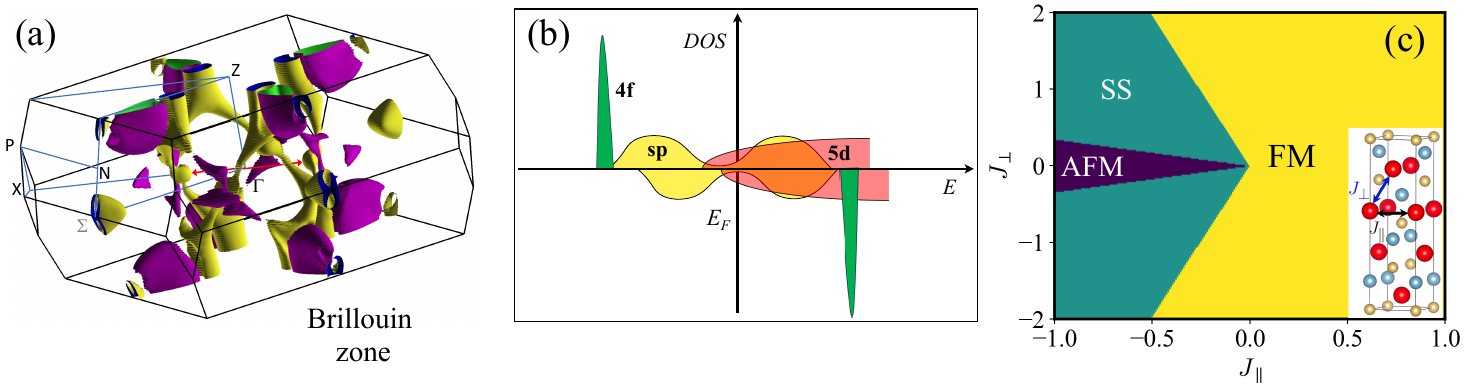}
\vspace{-0.7cm}
\caption{(a) The complex FS of semi-metallic SmAlSi, with the high-symmetry points and Brillouin zone 
indicated. The red arrow indicates the non-trivial nesting vector at the FS discussed in Ref.~\cite{Yao2023}. (b) Schematic representation of the low semimetallic density of states (DOS) at $E_\text{F}$ displaying the $sp$-band and the 
hybridization with spin-polarized $d$-electrons. (c) A simplified two-parameter phase diagram 
incorporating nearest neighbor intra-layer $(J_{\parallel})$ and inter-layer $(J_{\perp})$ interactions (see inset in Fig.~\ref{fig_all_REAlSib}c), 
illustrating the emergence of spin-spiral (SS) order within each rare-earth layer under antiferromagnetic ($J_{\parallel}<0$) intra-layer exchange interactions. Phases with (anti)ferromagnetic order within each layer are labeled with FM (AFM).}
\label{fig_all_REAlSib}
\vspace{-0.5cm}
\end{figure*}
For PrAlSi, the maximum peak occurs at $\boldsymbol{q}_p=(0,0,0)$ and the ferromagnetic order within each layer
is favored as observed experimentally in Ref.~\cite{Yang2021}, while for the NdAlSi case, 
the maximum peak occurs at $\boldsymbol{q}_p=\frac{2\pi}{a} (q_p,q_p,0)$ with $q_p = 0.273$ in agreement with the experimental value of $q^\text{exp}_p= \frac{1}{3}+\delta$~\cite{gaudet2021weyl}. 
However, the small energy difference $\Delta E= 0.07$ meV  between the spiral and ferromagnetic states, $J(\boldsymbol{q}_p) - J(\boldsymbol{0})$, is not sufficient to overcome the magnetic anisotropy energy caused by the crystal field, 
suppressing the helical ordering at low temperature and enforcing a ferromagnetic order intra-layer~\cite{gaudet2021weyl}. 
For the case of SmAlSi, the maximum peak occurs once more at a finite-$q$ with 
$\boldsymbol{q}_p=\frac{2\pi}{a} (q_p,q_p,0)$ and $q_p = 0.283$, 
very close to the experimental value $q^\text{exp}_p=0.33$~\cite{Yao2023}. 
Compared to the NdAlSi case, the energy difference here is much larger, $\Delta E= 0.36$ meV, and together with the reduction of the crystal field effects owing to the lanthanide 
lattice contraction of heavier $4f$ elements (see discussion below), this results in a persisting incommensurate order at low temperatures~\cite{Yao2023}. At these temperatures
quantum effects play a role in determining the transition temperatures. Nonetheless, estimating
$T_\text{N}$ using a mean-field, classical spin prescription, $T_\text{N}={J(\boldsymbol{q}_p)}/{3k_\text{B}}$, with 
$k_\text{B}$ being the Boltzmann constant, we find $T_\text{N}$  $\simeq 25$~K, which 
is of the same order of magnitude as the experimentally measured $T_\text{N}$~\cite{Yao2023}. 

To determine the dominant exchange contribution in $R$AlSi, we examine Fig.~\ref{fig_all_REAlSia}(b), where the black curves represent the isotropic exchange interactions $J_{ij}$ as functions of inter-atomic distance. The full (dotted) line represents short-range, antiferromagnetic intra-layer (inter-layer) interactions indicating a SX mechanism~\cite{Anderson1950} over a weak RKKY exchange. Despite the coexistence with RKKY-like interactions, the scarcity of electronic states near $E_\text{F}$ and the threadbare FS (Fig.~\ref{fig_all_REAlSib}(a)) favor the dominance of SX interactions.

Figure~\ref{fig_all_REAlSib}(a) displays the FS of the Gd-prototype (with the SmAlSi 
lattice constant) obtained using the FLEUR code~\cite{fleur} (see supplementary note 4). The FS occupies only a small portion of the Brillouin zone owing to the low density of states near the Fermi energy ($E_\text{F}$) \textit{i.e.}\ the semi-metallic 
nature of SmAlSi. The suggested nesting between the non-trivial Fermi pockets is 
indicated by a red arrow in Fig.~\ref{fig_all_REAlSib}(a). However, the FS 
sheets are not parallel and exhibit a three-dimensional dispersion, which does not 
fulfill the nesting condition required to stabilize incommensurate ordering. This 
observation aligns with recent findings in Ref.~\cite{zhang2023kramers}, where the 
computation of the Lindhard susceptibility, based on the FS, does not 
indicate a finite-$q$ helical ordering.

To further elaborate on this balance between RKKY and SX interactions, we analyze a related but metallic compound GdAl$_2$ with the same lattice constants (SmAlSi). Substituting Si with Al removes 
one electron from the system, shifting $E_\text{F}$ below the semi-metallic gap (see supplementary note 5). The resulting magnetic interactions are shown in Fig.~\ref{fig_all_REAlSia}(b). In contrast to GdAlSi, they 
exhibit an oscillatory long-range behavior, indicating that the 
RKKY interactions dominates over SX in this metallic regime,
favoring a spiroidal state (see inset of Fig.~\ref{fig_all_REAlSia}(b)). The SX mechanism in $R$AlSi 
compounds can be understood through an analysis of the density of states, as depicted 
schematically in Fig.~\ref{fig_all_REAlSib}(b). In these compounds, the $4f$ electrons 
induce a local Zeeman magnetic field, causing spin polarization of the rare-earth's $5d$-electrons. These 
induced $5d$ magnetic moments interact with the $p$ electrons
at different sites through the {non-magnetic} Si and Al atoms. The SX interactions can be described in terms of a charge-transfer model similar to the transition-metal
oxides one~\cite{gubanov2012magnetism}: $J_{ij}\propto -t^4_{pd}/\Delta^3$, where $t_{pd}$ 
represents the hopping integral between the $d$ and $p$ orbitals, and $\Delta$ is the 
charge-transfer energy. A comparable scenario arises in Gd monopnictides where SX 
competes with RKKY interactions~\cite{duan2006magnetic}. Lastly, while maintaining 
the same crystalline configuration but substituting Gd with Eu (EuAlSi), the $R$ valence $5d$ 
electrons are removed (the rare earth atoms are divalent rather than trivalent), 
leaving only $sp$-electrons that scatter off the localized $4f$ electrons. This ultimately 
leads to an RKKY interaction among the $4f$ moments (supplementary note 6), once 
again emphasizing the significant role of $5d$-electrons in generating the SX mechanism.

Although our analysis thus far indicates a dominant role for SX, we now
investigate the interplay between the Weyl points and the incommensurate 
helical ordering by considering two distinct cases: firstly excluding SOI, where the band structure contains nodal lines~\cite{Chang2018,zhang2023kramers},
 and secondly including SOI which introduces gaps in the nodal lines, and generates Weyl points at 
specific symmetry dictated positions in $k$-space~\cite{Chang2018}. Fig.~\ref{fig_all_REAlSia}(c)
depicts $J({q})$ of SmAlSi for these two cases - including (black curve) and excluding 
the SOI (green curve). While $J(\boldsymbol{q})$ differs slightly near $({q}_p, {q}_p, 0) $, the finite-$q$ peak structure remains and incommensurate order is still favored over ferromagnetism 
with an energy barrier $\Delta E= 0.10$~meV. This demonstrates that a stable helical 
order can form in the absence of Weyl points. Moreover, despite the presence of  Weyl nodes in the
electronic band structure of PrAlSi, $J(\boldsymbol{q})$ has the maximal value at $\boldsymbol{q}=(0,0,0)$ 
favoring a collinear order even without the inclusion of single-ion anisotropy 
(see Fig.~\ref{fig_all_REAlSia}(a)).
 
We now inspect the role of inversion-symmetry breaking on the magnetic order of 
SmAlSi. The centrosymmetry can be restored using a $50\%$-$50\%$ alloy of Al and 
Si species at each site, \textit{i.e.}\ Sm(Al$_{0.5}$Si$_{0.5}$)$_2$. This is achieved computationally  using the coherent potential 
approximation~\cite{Gyorffy1972}. The electronic band structure obtained for 
this centrosymmetric SmAlSi alloy is given in the supplementary note 7. Centrosymmetry
leads to the removal of the Weyl points (note that the DMI is suppressed as well) 
and the disorder smears out the electronic bands near the Fermi 
energy~\cite{ebert2011calculating}. The resulting $J(\boldsymbol{q})$, shown in Fig.~\ref{fig_all_REAlSia}(c) 
(blue curve), supports ferromagnetic order prevailing in the alloy ($\Delta E=-0.17$ meV), but 
the finite-$q$ peak persists indicating that the helical order does not originate 
from inversion symmetry breaking or the Weyl points. Lastly, inclusion of DMI 
(red curve) is found to have a minimal effect owing to its weak magnitude - while it breaks 
the $\pm {q}$ reciprocity reducing the four-fold degeneracy (inset Fig.~\ref{fig_all_REAlSia}(a)),
it does not alter the position or the magnitude of the finite-$q$ peak in $J(\boldsymbol{q})$ 
which is purely driven by isotropic exchange.

The $4f$ electrons' electronic configuration in $R$AlSi has an atomic-like behavior
in accordance with Hund's rules~\cite{Yao2023}, which in turn determines the shape of the 
$4f$ charge cloud. This charge is subjected to the crystal field (CF) originating from
the valence electrons and surrounding ions. Considering that the non-collinear order is 
driven by the isotropic exchange interactions and not crystal fields effects, 
the tetragonal uniaxial magnetic anisotropy constant $K_{i}$  is computed from fitting 
the classical CF energy differences~\cite{patrick2020spin} while rotating the $4f$ moment
from the $c$-axis to the $a$-axis. The crystal field parameters are obtained within the 
yttrium analogue approach~\cite{Patrick2019} (see supplementary note 8). Both PrAlSi 
and NdAlSi display an out-of-plane anisotropy, in agreement with the experimental 
observation~\cite{gaudet2021weyl}: PrAlSi has a high value of $K_{i}=2.593$ meV, 
while the NdAlSi constant is one order of magnitude smaller $K_{i}=0.218$ meV. 
On the other hand, SmAlSi prefers a canted easy anisotropy axis $\VEC{e}_{n}$ along 
the $(\theta_n,\phi_n)=(60^\circ,45^\circ)$ direction with respect to tetragonal 
basis vectors, with an anisotropy constant of $K_{i}= 0.13$ meV.

We extract the following minimal atomistic spin model in reciprocal space to understand 
the formation of a helical order in $R$AlSi from the short-ranged, antiferromagnetic 
interactions ($J<0$)
\begin{equation}
    \begin{cases}
      J^{\parallel}_{\mathrm{nn}}(\VEC{q}) = 2J_{\parallel}(\cos q_xa + \cos q_ya)\quad,\\
      J^{\perp}_{\mathrm{nm}}(\VEC{q}) = 2J_{\perp}(\cos (q_xa/2) + \cos (q_ya/2))\quad.
    \end{cases}       
\end{equation}
Hereby, we consider an isotropic Heisenberg model with nearest-neighbor intra-layer 
$(J_{\parallel})$ and inter-layer $(J_{\perp})$ interactions  as depicted in the inset 
of Fig.~\ref{fig_all_REAlSib}(c). We construct the phase diagram shown in Fig.~\ref{fig_all_REAlSib}(c)
by varying $J_{\parallel}$ and $J_{\perp}$, and identifying the in-plane $\VEC{q}$ which 
maximizes the eigenvalue $J(\VEC{q})$. Within each rare-earth layer three phases emerge, 
namely a ferromagnetic (FM), antiferromagnetic (AFM) and spin spiral (SS) phase. For 
$J_{\parallel}<0$, the AFM phase switches to a SS phase as the magnitude of $J_{\perp}$ 
increases. For SmAlSi, the $J_{\parallel}$ and $J_{\perp}$ parameters lie in the region
of the SS phase as can be seen from the first-principles results in Fig.~\ref{fig_all_REAlSia}(b). 
\begin{figure}
 \centering
\includegraphics[width=0.4\textwidth]{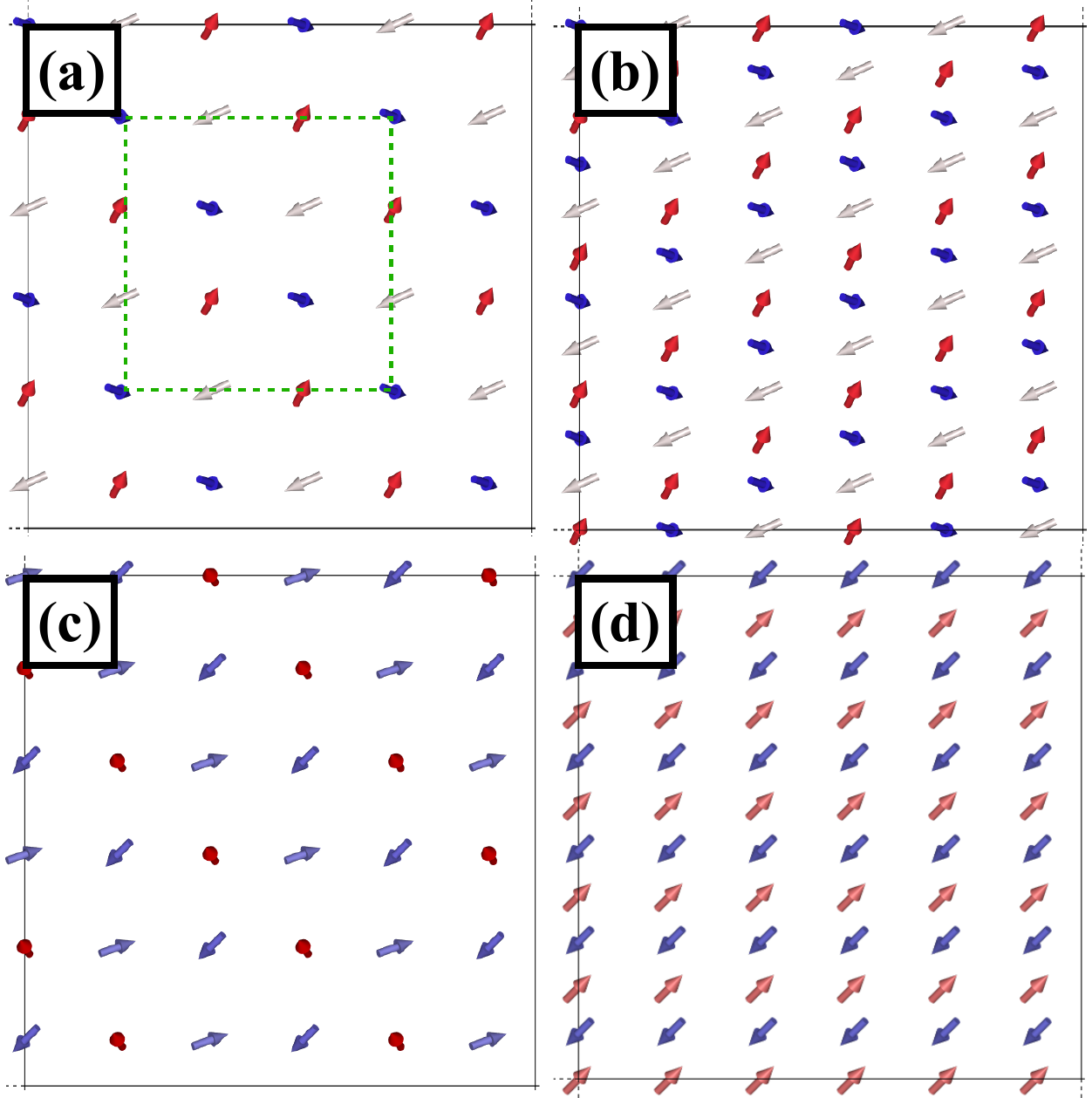}
\caption{(a) Helical magnetic structure of SmAlSi in the presence of frustrated isotropic 
exchange interactions and a canted magnetic anisotropy, the colors indicates the $m_z$ component, blue: 
$m_z=-1$, red: $m_z=1$, white: $m_z=0$). The green dashed square denotes magnetic unit cell. 
(b) Same as in (a), including sub-lattice two. (c) Same as in (a) including an external magnetic 
field ($B=4$\,Tesla) along the easy axis $(60^\circ, 45^\circ)$. (d) Stacked antiferromagnetic 
state along the $c$-axis with higher energy compared to (b).}
\vspace{-0.5cm}
\label{fig_magn_str}
\end{figure}

Now, for the case of SmAlSi, considering (short range) magnetic interactions $J_{ij}$ 
in real space up to $2.5$~$a$, with $K_{i}= 0.13$~meV, $\VEC{e}_n=(60^\circ,45^\circ)$, 
and neglecting the DMI, we minimize the Hamiltonian $\mathcal{H}_\text{m}$ \eqref{magn_ham} 
by solving the Landau-Lifshitz-Gilbert equation as implemented in the Spirit 
code~\cite{Gideon:2019}. The resulting helical order for the first sub-lattice is 
depicted in Fig.~\ref{fig_magn_str}(a) (top view), featuring a propagation vector 
$\VEC{q} = \frac{2\pi}{a} (q,q,0)$ with $q=0.33$ consistent with the maximum 
of $J(\VEC{q})$ for SmAlSi. Fig.~\ref{fig_magn_str}(b) illustrates the helimagnetic 
structure where both sub-lattices show a similar magnetic order but with 
a positional shift owing to the stacking along the $c$-axis. Besides the helical 
order, a solution slightly higher in energy ($\Delta E=0.05$ meV) features moments 
antiferromagnetically coupled along the $c$-axis, as depicted in Fig.~\ref{fig_magn_str}(d). 
The helical order can transform to this antiferromagnetic state under high-field 
conditions or with thermal fluctuations. To explore the emergence of 
non-collinear spin textures, we apply a magnetic field along the easy-axis~\cite{Bouaziz2022} 
$\VEC{e}_n$. The resulting state is shown in Fig.~\ref{fig_magn_str}(c). We observe 
a canting of the moments towards the $+z$ direction, but no skyrmion lattice phases 
can form owing to the short period of the magnetic structure (3$a$). The helical 
order remains the most stable one. 

Coming back to the above mentioned non-conventional contribution to the Hall 
signal observed experimentally for SmAlSi
when a magnetic field is applied~\cite{Yao2023}, 
we conjecture that this signal can be interpreted in 
terms of the recently introduced non-collinear Hall effect~\cite{Bouaziz2021}. 
This effect emerges from the interference between non-collinear magnetism and 
spin-orbit interactions in a non-centrosymmetric environment, without invoking 
the presence of magnetic skyrmions. Likely, this unconventional Hall signal 
originates from the scattering of Weyl fermions on the helical magnetic 
background.

In summary, we have examined three members from the $R$AlSi family, recently proposed as 
materials where Weyl-mediated RKKY interactions generate helical ordering. Our findings 
demonstrate that neither the Weyl points nor the RKKY interactions provide the predominant mechanism owing to the
presence of a low density of states near the Fermi energy in these semi-metals. Instead{,} we find a significant $p$-$d$ 
 antiferromagnetic SX contribution. The competition of  these isotropic SX
interactions between different atoms leads to a helical order with a short period of approximately 
three lattice constants, all without the assistance of DMI. Our \textit{ab initio} calculations reveal a strong magnetic anisotropy 
arising from crystal field effects for PrAlSi and NdAlSi, locking the moments into 
a collinear configuration. In contrast, the magnetic  anisotropy in SmAlSi is much lower, which allows the formation of a helical incommensurate order at low temperatures. Lastly, 
the short period of the spin spiral does not permit the emergence of a 
skyrmion lattice when external magnetic fields are applied. 
 
For future prospects, alloying Al and Si with other elements from the same family 
or applying strain may move the exchange parameters of the compound closer to the SS-FM phase boundaries 
of the phase diagram Fig.~\ref{fig_all_REAlSib}(b)  resulting in helical ordering with longer periods and lower magnetic fields needed to unwind the spiral. This 
could lead to the stabilization of skyrmions in the $R$AlSi materials family, providing 
an ideal platform to study the interplay between the topology of the magnetic 
texture in real space and that of Weyl fermions in reciprocal space.

\begin{acknowledgements}
We thank Carsten Timm, Collin L. Broholm, Nikolai Kiselev, and 
Andy Knoll for fruitful discussions. 
J.B.\ and S.B.\ acknowledges financial support from the 
European Research Council (ERC) under the European Union's Horizon 
2020 research and innovation program (Grant No.\ 856538, project 
``3D MAGiC''). S.B.\ acknowledges financial support from  Deutsche Forschungsgemeinschaft 
(DFG) through CRC 1238 (Project No.\ C01). J.B.S.\ acknowledges support from UK EPSRC Grant 
No.\ EP/M028941/1. J.B., S.B., and G.B.\  acknowledge 
computing time granted by the JARA-CSD and VSR Resource Allocation 
Board provided on the supercomputers CLAIX at RWTH Aachen University 
and JURECA at J\"ulich Supercomputer Centre under grants nos.\ jara0219 and jiff13. 
\end{acknowledgements}

\bibliography{bibliography.bib}

\begin{thebibliography}{45}%
\makeatletter
\providecommand \@ifxundefined [1]{%
 \@ifx{#1\undefined}
}%
\providecommand \@ifnum [1]{%
 \ifnum #1\expandafter \@firstoftwo
 \else \expandafter \@secondoftwo
 \fi
}%
\providecommand \@ifx [1]{%
 \ifx #1\expandafter \@firstoftwo
 \else \expandafter \@secondoftwo
 \fi
}%
\providecommand \natexlab [1]{#1}%
\providecommand \enquote  [1]{``#1''}%
\providecommand \bibnamefont  [1]{#1}%
\providecommand \bibfnamefont [1]{#1}%
\providecommand \citenamefont [1]{#1}%
\providecommand \href@noop [0]{\@secondoftwo}%
\providecommand \href [0]{\begingroup \@sanitize@url \@href}%
\providecommand \@href[1]{\@@startlink{#1}\@@href}%
\providecommand \@@href[1]{\endgroup#1\@@endlink}%
\providecommand \@sanitize@url [0]{\catcode `\\12\catcode `\$12\catcode
  `\&12\catcode `\#12\catcode `\^12\catcode `\_12\catcode `\%12\relax}%
\providecommand \@@startlink[1]{}%
\providecommand \@@endlink[0]{}%
\providecommand \url  [0]{\begingroup\@sanitize@url \@url }%
\providecommand \@url [1]{\endgroup\@href {#1}{\urlprefix }}%
\providecommand \urlprefix  [0]{URL }%
\providecommand \Eprint [0]{\href }%
\providecommand \doibase [0]{http://dx.doi.org/}%
\providecommand \selectlanguage [0]{\@gobble}%
\providecommand \bibinfo  [0]{\@secondoftwo}%
\providecommand \bibfield  [0]{\@secondoftwo}%
\providecommand \translation [1]{[#1]}%
\providecommand \BibitemOpen [0]{}%
\providecommand \bibitemStop [0]{}%
\providecommand \bibitemNoStop [0]{.\EOS\space}%
\providecommand \EOS [0]{\spacefactor3000\relax}%
\providecommand \BibitemShut  [1]{\csname bibitem#1\endcsname}%
\let\auto@bib@innerbib\@empty
\bibitem [{\citenamefont {Kumar}\ \emph {et~al.}(2020)\citenamefont {Kumar},
  \citenamefont {Guin}, \citenamefont {Manna}, \citenamefont {Shekhar},\ and\
  \citenamefont {Felser}}]{kumar2020topological}%
  \BibitemOpen
  \bibfield  {author} {\bibinfo {author} {\bibfnamefont {Nitesh}\ \bibnamefont
  {Kumar}}, \bibinfo {author} {\bibfnamefont {Satya~N}\ \bibnamefont {Guin}},
  \bibinfo {author} {\bibfnamefont {Kaustuv}\ \bibnamefont {Manna}}, \bibinfo
  {author} {\bibfnamefont {Chandra}\ \bibnamefont {Shekhar}}, \ and\ \bibinfo
  {author} {\bibfnamefont {Claudia}\ \bibnamefont {Felser}},\ }\bibfield
  {title} {\enquote {\bibinfo {title} {Topological quantum materials from the
  viewpoint of chemistry},}\ }\href@noop {} {\bibfield  {journal} {\bibinfo
  {journal} {Chemical Reviews}\ }\textbf {\bibinfo {volume} {121}},\ \bibinfo
  {pages} {2780--2815} (\bibinfo {year} {2020})}\BibitemShut {NoStop}%
\bibitem [{\citenamefont {Hasan}\ \emph {et~al.}(2021)\citenamefont {Hasan},
  \citenamefont {Chang}, \citenamefont {Belopolski}, \citenamefont {Bian},
  \citenamefont {Xu},\ and\ \citenamefont {Yin}}]{hasan2021weyl}%
  \BibitemOpen
  \bibfield  {author} {\bibinfo {author} {\bibfnamefont {M~Zahid}\ \bibnamefont
  {Hasan}}, \bibinfo {author} {\bibfnamefont {Guoqing}\ \bibnamefont {Chang}},
  \bibinfo {author} {\bibfnamefont {Ilya}\ \bibnamefont {Belopolski}}, \bibinfo
  {author} {\bibfnamefont {Guang}\ \bibnamefont {Bian}}, \bibinfo {author}
  {\bibfnamefont {Su-Yang}\ \bibnamefont {Xu}}, \ and\ \bibinfo {author}
  {\bibfnamefont {Jia-Xin}\ \bibnamefont {Yin}},\ }\bibfield  {title} {\enquote
  {\bibinfo {title} {{Weyl, Dirac and high-fold chiral fermions in topological
  quantum matter}},}\ }\href@noop {} {\bibfield  {journal} {\bibinfo  {journal}
  {Nature Reviews Materials}\ }\textbf {\bibinfo {volume} {6}},\ \bibinfo
  {pages} {784--803} (\bibinfo {year} {2021})}\BibitemShut {NoStop}%
\bibitem [{\citenamefont {Yan}\ and\ \citenamefont
  {Felser}(2017)}]{yan2017topological}%
  \BibitemOpen
  \bibfield  {author} {\bibinfo {author} {\bibfnamefont {Binghai}\ \bibnamefont
  {Yan}}\ and\ \bibinfo {author} {\bibfnamefont {Claudia}\ \bibnamefont
  {Felser}},\ }\bibfield  {title} {\enquote {\bibinfo {title} {{Topological
  materials: Weyl semimetals}},}\ }\href@noop {} {\bibfield  {journal}
  {\bibinfo  {journal} {Annual Review of Condensed Matter Physics}\ }\textbf
  {\bibinfo {volume} {8}},\ \bibinfo {pages} {337--354} (\bibinfo {year}
  {2017})}\BibitemShut {NoStop}%
\bibitem [{\citenamefont {Destraz}\ \emph {et~al.}(2020)\citenamefont
  {Destraz}, \citenamefont {Das}, \citenamefont {Tsirkin}, \citenamefont {Xu},
  \citenamefont {Neupert}, \citenamefont {Chang}, \citenamefont {Schilling},
  \citenamefont {Grushin}, \citenamefont {Kohlbrecher}, \citenamefont {Keller}
  \emph {et~al.}}]{destraz2020magnetism}%
  \BibitemOpen
  \bibfield  {author} {\bibinfo {author} {\bibfnamefont {Daniel}\ \bibnamefont
  {Destraz}}, \bibinfo {author} {\bibfnamefont {Lakshmi}\ \bibnamefont {Das}},
  \bibinfo {author} {\bibfnamefont {Stepan~S}\ \bibnamefont {Tsirkin}},
  \bibinfo {author} {\bibfnamefont {Yang}\ \bibnamefont {Xu}}, \bibinfo
  {author} {\bibfnamefont {Titus}\ \bibnamefont {Neupert}}, \bibinfo {author}
  {\bibfnamefont {J}~\bibnamefont {Chang}}, \bibinfo {author} {\bibfnamefont
  {A}~\bibnamefont {Schilling}}, \bibinfo {author} {\bibfnamefont {Adolfo~G}\
  \bibnamefont {Grushin}}, \bibinfo {author} {\bibfnamefont {Joachim}\
  \bibnamefont {Kohlbrecher}}, \bibinfo {author} {\bibfnamefont {Lukas}\
  \bibnamefont {Keller}},  \emph {et~al.},\ }\bibfield  {title} {\enquote
  {\bibinfo {title} {{Magnetism and anomalous transport in the Weyl semimetal
  PrAlGe: possible route to axial gauge fields}},}\ }\href@noop {} {\bibfield
  {journal} {\bibinfo  {journal} {npj Quantum Materials}\ }\textbf {\bibinfo
  {volume} {5}},\ \bibinfo {pages} {5} (\bibinfo {year} {2020})}\BibitemShut
  {NoStop}%
\bibitem [{\citenamefont {Huang}\ \emph {et~al.}(2015)\citenamefont {Huang},
  \citenamefont {Zhao}, \citenamefont {Long}, \citenamefont {Wang},
  \citenamefont {Chen}, \citenamefont {Yang}, \citenamefont {Liang},
  \citenamefont {Xue}, \citenamefont {Weng}, \citenamefont {Fang} \emph
  {et~al.}}]{huang2015observation}%
  \BibitemOpen
  \bibfield  {author} {\bibinfo {author} {\bibfnamefont {Xiaochun}\
  \bibnamefont {Huang}}, \bibinfo {author} {\bibfnamefont {Lingxiao}\
  \bibnamefont {Zhao}}, \bibinfo {author} {\bibfnamefont {Yujia}\ \bibnamefont
  {Long}}, \bibinfo {author} {\bibfnamefont {Peipei}\ \bibnamefont {Wang}},
  \bibinfo {author} {\bibfnamefont {Dong}\ \bibnamefont {Chen}}, \bibinfo
  {author} {\bibfnamefont {Zhanhai}\ \bibnamefont {Yang}}, \bibinfo {author}
  {\bibfnamefont {Hui}\ \bibnamefont {Liang}}, \bibinfo {author} {\bibfnamefont
  {Mianqi}\ \bibnamefont {Xue}}, \bibinfo {author} {\bibfnamefont {Hongming}\
  \bibnamefont {Weng}}, \bibinfo {author} {\bibfnamefont {Zhong}\ \bibnamefont
  {Fang}},  \emph {et~al.},\ }\bibfield  {title} {\enquote {\bibinfo {title}
  {{Observation of the chiral-anomaly-induced negative magnetoresistance in 3D
  Weyl semimetal TaAs}},}\ }\href@noop {} {\bibfield  {journal} {\bibinfo
  {journal} {Physical Review X}\ }\textbf {\bibinfo {volume} {5}},\ \bibinfo
  {pages} {031023} (\bibinfo {year} {2015})}\BibitemShut {NoStop}%
\bibitem [{\citenamefont {Knoll}\ \emph {et~al.}(2020)\citenamefont {Knoll},
  \citenamefont {Timm},\ and\ \citenamefont {Meng}}]{Knoll2020}%
  \BibitemOpen
  \bibfield  {author} {\bibinfo {author} {\bibfnamefont {Andy}\ \bibnamefont
  {Knoll}}, \bibinfo {author} {\bibfnamefont {Carsten}\ \bibnamefont {Timm}}, \
  and\ \bibinfo {author} {\bibfnamefont {Tobias}\ \bibnamefont {Meng}},\
  }\bibfield  {title} {\enquote {\bibinfo {title} {{Negative longitudinal
  magnetoconductance at weak fields in Weyl semimetals}},}\ }\href@noop {}
  {\bibfield  {journal} {\bibinfo  {journal} {Phys. Rev. B}\ }\textbf {\bibinfo
  {volume} {101}},\ \bibinfo {pages} {201402} (\bibinfo {year}
  {2020})}\BibitemShut {NoStop}%
\bibitem [{\citenamefont {Nagaosa}\ \emph {et~al.}(2020)\citenamefont
  {Nagaosa}, \citenamefont {Morimoto},\ and\ \citenamefont
  {Tokura}}]{nagaosa2020transport}%
  \BibitemOpen
  \bibfield  {author} {\bibinfo {author} {\bibfnamefont {Naoto}\ \bibnamefont
  {Nagaosa}}, \bibinfo {author} {\bibfnamefont {Takahiro}\ \bibnamefont
  {Morimoto}}, \ and\ \bibinfo {author} {\bibfnamefont {Yoshinori}\
  \bibnamefont {Tokura}},\ }\bibfield  {title} {\enquote {\bibinfo {title}
  {{Transport, magnetic and optical properties of Weyl materials}},}\
  }\href@noop {} {\bibfield  {journal} {\bibinfo  {journal} {Nature Reviews
  Materials}\ }\textbf {\bibinfo {volume} {5}},\ \bibinfo {pages} {621--636}
  (\bibinfo {year} {2020})}\BibitemShut {NoStop}%
\bibitem [{\citenamefont {Knoll}\ and\ \citenamefont {Timm}(2022)}]{Knoll2022}%
  \BibitemOpen
  \bibfield  {author} {\bibinfo {author} {\bibfnamefont {Andy}\ \bibnamefont
  {Knoll}}\ and\ \bibinfo {author} {\bibfnamefont {Carsten}\ \bibnamefont
  {Timm}},\ }\bibfield  {title} {\enquote {\bibinfo {title} {Classification of
  {W}eyl points and nodal lines based on magnetic point groups for
  spin-$\frac{1}{2}$ quasiparticles},}\ }\href@noop {} {\bibfield  {journal}
  {\bibinfo  {journal} {Phys. Rev. B}\ }\textbf {\bibinfo {volume} {105}},\
  \bibinfo {pages} {115109} (\bibinfo {year} {2022})}\BibitemShut {NoStop}%
\bibitem [{\citenamefont {Dzyaloshinsky}(1958)}]{Dzyaloshinsky:1958}%
  \BibitemOpen
  \bibfield  {author} {\bibinfo {author} {\bibfnamefont {I.}~\bibnamefont
  {Dzyaloshinsky}},\ }\bibfield  {title} {\enquote {\bibinfo {title} {{A
  thermodynamic theory of “weak” ferromagnetism of antiferromagnetics}},}\
  }\href@noop {} {\bibfield  {journal} {\bibinfo  {journal} {Journal of Physics
  and Chemistry of Solids}\ }\textbf {\bibinfo {volume} {4}},\ \bibinfo {pages}
  {241--255} (\bibinfo {year} {1958})}\BibitemShut {NoStop}%
\bibitem [{\citenamefont {Moriya}(1960)}]{Moriya:1960}%
  \BibitemOpen
  \bibfield  {author} {\bibinfo {author} {\bibfnamefont {T.}~\bibnamefont
  {Moriya}},\ }\bibfield  {title} {\enquote {\bibinfo {title} {{Anisotropic
  superexchange interaction and weak ferromagnetism}},}\ }\href@noop {}
  {\bibfield  {journal} {\bibinfo  {journal} {Physical Review}\ }\textbf
  {\bibinfo {volume} {120}},\ \bibinfo {pages} {91} (\bibinfo {year}
  {1960})}\BibitemShut {NoStop}%
\bibitem [{\citenamefont {Dzyaloshinsky}(1964)}]{Dzyaloshinsky:1964}%
  \BibitemOpen
  \bibfield  {author} {\bibinfo {author} {\bibfnamefont {I.}~\bibnamefont
  {Dzyaloshinsky}},\ }\bibfield  {title} {\enquote {\bibinfo {title} {{Theory
  of helicoidal structures in antiferromagnets}},}\ }\href@noop {} {\bibfield
  {journal} {\bibinfo  {journal} {Journal of Exprimenta and Theoreitcal Physics
  (U.S.S.R.)}\ }\textbf {\bibinfo {volume} {74}},\ \bibinfo {pages} {992--1002}
  (\bibinfo {year} {1964})}\BibitemShut {NoStop}%
\bibitem [{\citenamefont {Bogdanov}\ and\ \citenamefont
  {Yablonski}(1989)}]{Bogdanov:1989}%
  \BibitemOpen
  \bibfield  {author} {\bibinfo {author} {\bibfnamefont {A.}~\bibnamefont
  {Bogdanov}}\ and\ \bibinfo {author} {\bibfnamefont {D.}~\bibnamefont
  {Yablonski}},\ }\bibfield  {title} {\enquote {\bibinfo {title}
  {{Thermodynamicallystable "vortices" in magnetically ordered crystals. The
  mixed state of magnets}},}\ }\href@noop {} {\bibfield  {journal} {\bibinfo
  {journal} {Zh. Eksp. Teor. Fiz.}\ }\textbf {\bibinfo {volume} {95}},\
  \bibinfo {pages} {178} (\bibinfo {year} {1989})}\BibitemShut {NoStop}%
\bibitem [{\citenamefont {Gaudet}\ \emph {et~al.}(2021)\citenamefont {Gaudet},
  \citenamefont {Yang}, \citenamefont {Baidya}, \citenamefont {Lu},
  \citenamefont {Xu}, \citenamefont {Zhao}, \citenamefont {Rodriguez-Rivera},
  \citenamefont {Hoffmann}, \citenamefont {Graf}, \citenamefont {Torchinsky}
  \emph {et~al.}}]{gaudet2021weyl}%
  \BibitemOpen
  \bibfield  {author} {\bibinfo {author} {\bibfnamefont {Jonathan}\
  \bibnamefont {Gaudet}}, \bibinfo {author} {\bibfnamefont {Hung-Yu}\
  \bibnamefont {Yang}}, \bibinfo {author} {\bibfnamefont {Santu}\ \bibnamefont
  {Baidya}}, \bibinfo {author} {\bibfnamefont {Baozhu}\ \bibnamefont {Lu}},
  \bibinfo {author} {\bibfnamefont {Guangyong}\ \bibnamefont {Xu}}, \bibinfo
  {author} {\bibfnamefont {Yang}\ \bibnamefont {Zhao}}, \bibinfo {author}
  {\bibfnamefont {Jose~A}\ \bibnamefont {Rodriguez-Rivera}}, \bibinfo {author}
  {\bibfnamefont {Christina~M}\ \bibnamefont {Hoffmann}}, \bibinfo {author}
  {\bibfnamefont {David~E}\ \bibnamefont {Graf}}, \bibinfo {author}
  {\bibfnamefont {Darius~H}\ \bibnamefont {Torchinsky}},  \emph {et~al.},\
  }\bibfield  {title} {\enquote {\bibinfo {title} {{Weyl-mediated helical
  magnetism in NdAlSi}},}\ }\href@noop {} {\bibfield  {journal} {\bibinfo
  {journal} {Nature Materials}\ }\textbf {\bibinfo {volume} {20}},\ \bibinfo
  {pages} {1650--1656} (\bibinfo {year} {2021})}\BibitemShut {NoStop}%
\bibitem [{\citenamefont {Yao}\ \emph {et~al.}(2023{\natexlab{a}})\citenamefont
  {Yao}, \citenamefont {Gaudet}, \citenamefont {Verma}, \citenamefont {Graf},
  \citenamefont {Yang}, \citenamefont {Bahrami}, \citenamefont {Zhang},
  \citenamefont {Aczel}, \citenamefont {Subedi}, \citenamefont {Torchinsky}
  \emph {et~al.}}]{yao2023large}%
  \BibitemOpen
  \bibfield  {author} {\bibinfo {author} {\bibfnamefont {Xiaohan}\ \bibnamefont
  {Yao}}, \bibinfo {author} {\bibfnamefont {Jonathan}\ \bibnamefont {Gaudet}},
  \bibinfo {author} {\bibfnamefont {Rahul}\ \bibnamefont {Verma}}, \bibinfo
  {author} {\bibfnamefont {David~E}\ \bibnamefont {Graf}}, \bibinfo {author}
  {\bibfnamefont {Hung-Yu}\ \bibnamefont {Yang}}, \bibinfo {author}
  {\bibfnamefont {Faranak}\ \bibnamefont {Bahrami}}, \bibinfo {author}
  {\bibfnamefont {Ruiqi}\ \bibnamefont {Zhang}}, \bibinfo {author}
  {\bibfnamefont {Adam~A}\ \bibnamefont {Aczel}}, \bibinfo {author}
  {\bibfnamefont {Sujan}\ \bibnamefont {Subedi}}, \bibinfo {author}
  {\bibfnamefont {Darius~H}\ \bibnamefont {Torchinsky}},  \emph {et~al.},\
  }\bibfield  {title} {\enquote {\bibinfo {title} {{Large Topological Hall
  Effect and Spiral Magnetic Order in the Weyl Semimetal SmAlSi}},}\
  }\href@noop {} {\bibfield  {journal} {\bibinfo  {journal} {Physical Review
  X}\ }\textbf {\bibinfo {volume} {13}},\ \bibinfo {pages} {011035} (\bibinfo
  {year} {2023}{\natexlab{a}})}\BibitemShut {NoStop}%
\bibitem [{\citenamefont {Lyu}\ \emph {et~al.}(2020)\citenamefont {Lyu},
  \citenamefont {Xiang}, \citenamefont {Mi}, \citenamefont {Zhao},
  \citenamefont {Wang}, \citenamefont {Liu}, \citenamefont {Chen},
  \citenamefont {Ren}, \citenamefont {Li},\ and\ \citenamefont
  {Sun}}]{lyu2020nonsaturating}%
  \BibitemOpen
  \bibfield  {author} {\bibinfo {author} {\bibfnamefont {Meng}\ \bibnamefont
  {Lyu}}, \bibinfo {author} {\bibfnamefont {Junsen}\ \bibnamefont {Xiang}},
  \bibinfo {author} {\bibfnamefont {Zhenyu}\ \bibnamefont {Mi}}, \bibinfo
  {author} {\bibfnamefont {Hengcan}\ \bibnamefont {Zhao}}, \bibinfo {author}
  {\bibfnamefont {Zhen}\ \bibnamefont {Wang}}, \bibinfo {author} {\bibfnamefont
  {Enke}\ \bibnamefont {Liu}}, \bibinfo {author} {\bibfnamefont {Genfu}\
  \bibnamefont {Chen}}, \bibinfo {author} {\bibfnamefont {Zhian}\ \bibnamefont
  {Ren}}, \bibinfo {author} {\bibfnamefont {Gang}\ \bibnamefont {Li}}, \ and\
  \bibinfo {author} {\bibfnamefont {Peijie}\ \bibnamefont {Sun}},\ }\bibfield
  {title} {\enquote {\bibinfo {title} {{Nonsaturating magnetoresistance,
  anomalous Hall effect, and magnetic quantum oscillations in the ferromagnetic
  semimetal PrAlSi}},}\ }\href@noop {} {\bibfield  {journal} {\bibinfo
  {journal} {Physical Review B}\ }\textbf {\bibinfo {volume} {102}},\ \bibinfo
  {pages} {085143} (\bibinfo {year} {2020})}\BibitemShut {NoStop}%
\bibitem [{\citenamefont {Puphal}\ \emph {et~al.}(2020)\citenamefont {Puphal},
  \citenamefont {Pomjakushin}, \citenamefont {Kanazawa}, \citenamefont
  {Ukleev}, \citenamefont {Gawryluk}, \citenamefont {Ma}, \citenamefont
  {Naamneh}, \citenamefont {Plumb}, \citenamefont {Keller}, \citenamefont
  {Cubitt} \emph {et~al.}}]{puphal2020topological}%
  \BibitemOpen
  \bibfield  {author} {\bibinfo {author} {\bibfnamefont {Pascal}\ \bibnamefont
  {Puphal}}, \bibinfo {author} {\bibfnamefont {Vladimir}\ \bibnamefont
  {Pomjakushin}}, \bibinfo {author} {\bibfnamefont {Naoya}\ \bibnamefont
  {Kanazawa}}, \bibinfo {author} {\bibfnamefont {Victor}\ \bibnamefont
  {Ukleev}}, \bibinfo {author} {\bibfnamefont {Dariusz~J}\ \bibnamefont
  {Gawryluk}}, \bibinfo {author} {\bibfnamefont {Junzhang}\ \bibnamefont {Ma}},
  \bibinfo {author} {\bibfnamefont {Muntaser}\ \bibnamefont {Naamneh}},
  \bibinfo {author} {\bibfnamefont {Nicholas~C}\ \bibnamefont {Plumb}},
  \bibinfo {author} {\bibfnamefont {Lukas}\ \bibnamefont {Keller}}, \bibinfo
  {author} {\bibfnamefont {Robert}\ \bibnamefont {Cubitt}},  \emph {et~al.},\
  }\bibfield  {title} {\enquote {\bibinfo {title} {{Topological magnetic phase
  in the candidate Weyl semimetal CeAlGe}},}\ }\href@noop {} {\bibfield
  {journal} {\bibinfo  {journal} {Physical Review Letters}\ }\textbf {\bibinfo
  {volume} {124}},\ \bibinfo {pages} {017202} (\bibinfo {year}
  {2020})}\BibitemShut {NoStop}%
\bibitem [{\citenamefont {Yang}\ \emph
  {et~al.}(2021{\natexlab{a}})\citenamefont {Yang}, \citenamefont {Singh},
  \citenamefont {Gaudet}, \citenamefont {Lu}, \citenamefont {Huang},
  \citenamefont {Chiu}, \citenamefont {Huang}, \citenamefont {Wang},
  \citenamefont {Bahrami}, \citenamefont {Xu} \emph
  {et~al.}}]{yang2021noncollinear}%
  \BibitemOpen
  \bibfield  {author} {\bibinfo {author} {\bibfnamefont {Hung-Yu}\ \bibnamefont
  {Yang}}, \bibinfo {author} {\bibfnamefont {Bahadur}\ \bibnamefont {Singh}},
  \bibinfo {author} {\bibfnamefont {Jonathan}\ \bibnamefont {Gaudet}}, \bibinfo
  {author} {\bibfnamefont {Baozhu}\ \bibnamefont {Lu}}, \bibinfo {author}
  {\bibfnamefont {Cheng-Yi}\ \bibnamefont {Huang}}, \bibinfo {author}
  {\bibfnamefont {Wei-Chi}\ \bibnamefont {Chiu}}, \bibinfo {author}
  {\bibfnamefont {Shin-Ming}\ \bibnamefont {Huang}}, \bibinfo {author}
  {\bibfnamefont {Baokai}\ \bibnamefont {Wang}}, \bibinfo {author}
  {\bibfnamefont {Faranak}\ \bibnamefont {Bahrami}}, \bibinfo {author}
  {\bibfnamefont {Bochao}\ \bibnamefont {Xu}},  \emph {et~al.},\ }\bibfield
  {title} {\enquote {\bibinfo {title} {{Noncollinear ferromagnetic Weyl
  semimetal with anisotropic anomalous Hall effect}},}\ }\href@noop {}
  {\bibfield  {journal} {\bibinfo  {journal} {Physical Review B}\ }\textbf
  {\bibinfo {volume} {103}},\ \bibinfo {pages} {115143} (\bibinfo {year}
  {2021}{\natexlab{a}})}\BibitemShut {NoStop}%
\bibitem [{\citenamefont {Jensen}\ and\ \citenamefont
  {Mackintosh}(1991)}]{mackintosh1991rare}%
  \BibitemOpen
  \bibfield  {author} {\bibinfo {author} {\bibfnamefont {J.}~\bibnamefont
  {Jensen}}\ and\ \bibinfo {author} {\bibfnamefont {Allan~R.}\ \bibnamefont
  {Mackintosh}},\ }\href@noop {} {\emph {\bibinfo {title} {Rare earth
  magnetism: structures and excitations}}}\ (\bibinfo  {publisher} {Clarendon
  Oxford},\ \bibinfo {year} {1991})\BibitemShut {NoStop}%
\bibitem [{\citenamefont {Neubauer}\ \emph {et~al.}(2009)\citenamefont
  {Neubauer}, \citenamefont {Pfleiderer}, \citenamefont {Binz}, \citenamefont
  {Rosch}, \citenamefont {Ritz}, \citenamefont {Niklowitz},\ and\ \citenamefont
  {B\"oni}}]{Neubauer2009}%
  \BibitemOpen
  \bibfield  {author} {\bibinfo {author} {\bibfnamefont {A.}~\bibnamefont
  {Neubauer}}, \bibinfo {author} {\bibfnamefont {C.}~\bibnamefont
  {Pfleiderer}}, \bibinfo {author} {\bibfnamefont {B.}~\bibnamefont {Binz}},
  \bibinfo {author} {\bibfnamefont {A.}~\bibnamefont {Rosch}}, \bibinfo
  {author} {\bibfnamefont {R.}~\bibnamefont {Ritz}}, \bibinfo {author}
  {\bibfnamefont {P.~G.}\ \bibnamefont {Niklowitz}}, \ and\ \bibinfo {author}
  {\bibfnamefont {P.}~\bibnamefont {B\"oni}},\ }\bibfield  {title} {\enquote
  {\bibinfo {title} {{Topological Hall Effect in the $A$ Phase of MnSi}},}\
  }\href@noop {} {\bibfield  {journal} {\bibinfo  {journal} {Phys. Rev. Lett.}\
  }\textbf {\bibinfo {volume} {102}},\ \bibinfo {pages} {186602} (\bibinfo
  {year} {2009})}\BibitemShut {NoStop}%
\bibitem [{\citenamefont {Anderson}(1950)}]{Anderson1950}%
  \BibitemOpen
  \bibfield  {author} {\bibinfo {author} {\bibfnamefont {P.~W.}\ \bibnamefont
  {Anderson}},\ }\bibfield  {title} {\enquote {\bibinfo {title}
  {Antiferromagnetism. theory of superexchange interaction},}\ }\href@noop {}
  {\bibfield  {journal} {\bibinfo  {journal} {Phys. Rev.}\ }\textbf {\bibinfo
  {volume} {79}},\ \bibinfo {pages} {350--356} (\bibinfo {year}
  {1950})}\BibitemShut {NoStop}%
\bibitem [{\citenamefont {Solovyev}\ and\ \citenamefont
  {Terakura}(1999)}]{Solovyev1999}%
  \BibitemOpen
  \bibfield  {author} {\bibinfo {author} {\bibfnamefont {I.~V.}\ \bibnamefont
  {Solovyev}}\ and\ \bibinfo {author} {\bibfnamefont {K.}~\bibnamefont
  {Terakura}},\ }\bibfield  {title} {\enquote {\bibinfo {title} {Zone boundary
  softening of the spin-wave dispersion in doped ferromagnetic manganites},}\
  }\href@noop {} {\bibfield  {journal} {\bibinfo  {journal} {Phys. Rev. Lett.}\
  }\textbf {\bibinfo {volume} {82}},\ \bibinfo {pages} {2959--2962} (\bibinfo
  {year} {1999})}\BibitemShut {NoStop}%
\bibitem [{\citenamefont {Solovyev}(1999)}]{Solovyev1999_2}%
  \BibitemOpen
  \bibfield  {author} {\bibinfo {author} {\bibfnamefont {I.~V.}\ \bibnamefont
  {Solovyev}},\ }\bibfield  {title} {\enquote {\bibinfo {title} {Optimized
  effective potential for the extended hubbard model},}\ }\href@noop {}
  {\bibfield  {journal} {\bibinfo  {journal} {Phys. Rev. B}\ }\textbf {\bibinfo
  {volume} {60}},\ \bibinfo {pages} {8550--8558} (\bibinfo {year}
  {1999})}\BibitemShut {NoStop}%
\bibitem [{\citenamefont {Hughes}\ \emph {et~al.}(2007)\citenamefont {Hughes},
  \citenamefont {D{\"a}ne}, \citenamefont {Ernst}, \citenamefont {Hergert},
  \citenamefont {L{\"u}ders}, \citenamefont {Poulter}, \citenamefont
  {Staunton}, \citenamefont {Svane}, \citenamefont {Szotek},\ and\
  \citenamefont {Temmerman}}]{hughes2007}%
  \BibitemOpen
  \bibfield  {author} {\bibinfo {author} {\bibfnamefont {I.~D.}\ \bibnamefont
  {Hughes}}, \bibinfo {author} {\bibfnamefont {M.}~\bibnamefont {D{\"a}ne}},
  \bibinfo {author} {\bibfnamefont {A.}~\bibnamefont {Ernst}}, \bibinfo
  {author} {\bibfnamefont {W.}~\bibnamefont {Hergert}}, \bibinfo {author}
  {\bibfnamefont {M.}~\bibnamefont {L{\"u}ders}}, \bibinfo {author}
  {\bibfnamefont {J.}~\bibnamefont {Poulter}}, \bibinfo {author} {\bibfnamefont
  {J.~B.}\ \bibnamefont {Staunton}}, \bibinfo {author} {\bibfnamefont
  {A.}~\bibnamefont {Svane}}, \bibinfo {author} {\bibfnamefont
  {Z.}~\bibnamefont {Szotek}}, \ and\ \bibinfo {author} {\bibfnamefont {W.~M.}\
  \bibnamefont {Temmerman}},\ }\bibfield  {title} {\enquote {\bibinfo {title}
  {Lanthanide contraction and magnetism in the heavy rare earth elements},}\
  }\href@noop {} {\bibfield  {journal} {\bibinfo  {journal} {Nature}\ }\textbf
  {\bibinfo {volume} {446}},\ \bibinfo {pages} {650--653} (\bibinfo {year}
  {2007})}\BibitemShut {NoStop}%
\bibitem [{\citenamefont {Mendive-Tapia}\ and\ \citenamefont
  {Staunton}(2017)}]{Mendive2017}%
  \BibitemOpen
  \bibfield  {author} {\bibinfo {author} {\bibfnamefont {E.}~\bibnamefont
  {Mendive-Tapia}}\ and\ \bibinfo {author} {\bibfnamefont {J.~B.}\ \bibnamefont
  {Staunton}},\ }\bibfield  {title} {\enquote {\bibinfo {title} {Theory of
  magnetic ordering in the heavy rare earths: Ab initio electronic origin of
  pair- and four-spin interactions},}\ }\href@noop {} {\bibfield  {journal}
  {\bibinfo  {journal} {Phys. Rev. Lett.}\ }\textbf {\bibinfo {volume} {118}},\
  \bibinfo {pages} {197202} (\bibinfo {year} {2017})}\BibitemShut {NoStop}%
\bibitem [{\citenamefont {Bouaziz}\ \emph {et~al.}(2022)\citenamefont
  {Bouaziz}, \citenamefont {Mendive-Tapia}, \citenamefont {Bl\"ugel},\ and\
  \citenamefont {Staunton}}]{Bouaziz2022}%
  \BibitemOpen
  \bibfield  {author} {\bibinfo {author} {\bibfnamefont {Juba}\ \bibnamefont
  {Bouaziz}}, \bibinfo {author} {\bibfnamefont {Eduardo}\ \bibnamefont
  {Mendive-Tapia}}, \bibinfo {author} {\bibfnamefont {Stefan}\ \bibnamefont
  {Bl\"ugel}}, \ and\ \bibinfo {author} {\bibfnamefont {Julie~B.}\ \bibnamefont
  {Staunton}},\ }\bibfield  {title} {\enquote {\bibinfo {title} {{Fermi-Surface
  Origin of Skyrmion Lattices in Centrosymmetric Rare-Earth Intermetallics}},}\
  }\href@noop {} {\bibfield  {journal} {\bibinfo  {journal} {Phys. Rev. Lett.}\
  }\textbf {\bibinfo {volume} {128}},\ \bibinfo {pages} {157206} (\bibinfo
  {year} {2022})}\BibitemShut {NoStop}%
\bibitem [{\citenamefont {Patrick}\ and\ \citenamefont
  {Staunton}(2019)}]{Patrick2019}%
  \BibitemOpen
  \bibfield  {author} {\bibinfo {author} {\bibfnamefont {Christopher~E}\
  \bibnamefont {Patrick}}\ and\ \bibinfo {author} {\bibfnamefont {Julie~B}\
  \bibnamefont {Staunton}},\ }\bibfield  {title} {\enquote {\bibinfo {title}
  {Crystal field coefficients for yttrium analogues of
  rare-earth/transition-metal magnets using density-functional theory in the
  projector-augmented wave formalism},}\ }\href@noop {} {\bibfield  {journal}
  {\bibinfo  {journal} {Journal of Physics: Condensed Matter}\ }\textbf
  {\bibinfo {volume} {31}},\ \bibinfo {pages} {305901} (\bibinfo {year}
  {2019})}\BibitemShut {NoStop}%
\bibitem [{\citenamefont {Dudarev}\ \emph {et~al.}(2019)\citenamefont
  {Dudarev}, \citenamefont {Liu}, \citenamefont {Andersson}, \citenamefont
  {Stanek}, \citenamefont {Ozaki},\ and\ \citenamefont
  {Franchini}}]{Dudarev2019}%
  \BibitemOpen
  \bibfield  {author} {\bibinfo {author} {\bibfnamefont {S.~L.}\ \bibnamefont
  {Dudarev}}, \bibinfo {author} {\bibfnamefont {P.}~\bibnamefont {Liu}},
  \bibinfo {author} {\bibfnamefont {D.~A.}\ \bibnamefont {Andersson}}, \bibinfo
  {author} {\bibfnamefont {C.~R.}\ \bibnamefont {Stanek}}, \bibinfo {author}
  {\bibfnamefont {T.}~\bibnamefont {Ozaki}}, \ and\ \bibinfo {author}
  {\bibfnamefont {C.}~\bibnamefont {Franchini}},\ }\bibfield  {title} {\enquote
  {\bibinfo {title} {{Parametrization of $\mathrm{LSDA}+U$ for noncollinear
  magnetic configurations: Multipolar magnetism in ${\mathrm{UO}}_{2}$}},}\
  }\href@noop {} {\bibfield  {journal} {\bibinfo  {journal} {Phys. Rev.
  Mater.}\ }\textbf {\bibinfo {volume} {3}},\ \bibinfo {pages} {083802}
  (\bibinfo {year} {2019})}\BibitemShut {NoStop}%
\bibitem [{\citenamefont {Papanikolaou}\ \emph {et~al.}(2002)\citenamefont
  {Papanikolaou}, \citenamefont {Zeller},\ and\ \citenamefont
  {Dederichs}}]{papanikolaou:2002}%
  \BibitemOpen
  \bibfield  {author} {\bibinfo {author} {\bibfnamefont {N.}~\bibnamefont
  {Papanikolaou}}, \bibinfo {author} {\bibfnamefont {R.}~\bibnamefont
  {Zeller}}, \ and\ \bibinfo {author} {\bibfnamefont {P.~H.}\ \bibnamefont
  {Dederichs}},\ }\bibfield  {title} {\enquote {\bibinfo {title} {{Conceptual
  improvements of the KKR method}},}\ }\href@noop {} {\bibfield  {journal}
  {\bibinfo  {journal} {Journal of Physics: Condensed Matter}\ }\textbf
  {\bibinfo {volume} {14}},\ \bibinfo {pages} {2799} (\bibinfo {year}
  {2002})}\BibitemShut {NoStop}%
\bibitem [{\citenamefont {Bauer}(2014)}]{Bauer:2014}%
  \BibitemOpen
  \bibfield  {author} {\bibinfo {author} {\bibfnamefont {D.~S.~G.}\
  \bibnamefont {Bauer}},\ }\bibfield  {title} {\enquote {\bibinfo {title}
  {{Development of a relativistic full-potential first-principles multiple
  scattering {G}reen function method applied to complex magnetic textures of
  nanostructures at surfaces}},}\ }\href@noop {} {\bibfield  {journal}
  {\bibinfo  {journal} {Forschungszentrum J\"ulich}\ } (\bibinfo {year}
  {2014})}\BibitemShut {NoStop}%
\bibitem [{\citenamefont {Liechtenstein}\ \emph {et~al.}(1984)\citenamefont
  {Liechtenstein}, \citenamefont {Katsnelson},\ and\ \citenamefont
  {Gubanov}}]{liechtenstein1984}%
  \BibitemOpen
  \bibfield  {author} {\bibinfo {author} {\bibfnamefont {A.~I.}\ \bibnamefont
  {Liechtenstein}}, \bibinfo {author} {\bibfnamefont {M.~I.}\ \bibnamefont
  {Katsnelson}}, \ and\ \bibinfo {author} {\bibfnamefont {V.~A.}\ \bibnamefont
  {Gubanov}},\ }\bibfield  {title} {\enquote {\bibinfo {title} {Exchange
  interactions and spin-wave stiffness in ferromagnetic metals},}\ }\href@noop
  {} {\bibfield  {journal} {\bibinfo  {journal} {Journal of Physics F: Metal
  Physics}\ }\textbf {\bibinfo {volume} {14}},\ \bibinfo {pages} {L125}
  (\bibinfo {year} {1984})}\BibitemShut {NoStop}%
\bibitem [{\citenamefont {Ebert}\ and\ \citenamefont
  {Mankovsky}(2009)}]{Ebert:2009}%
  \BibitemOpen
  \bibfield  {author} {\bibinfo {author} {\bibfnamefont {H.}~\bibnamefont
  {Ebert}}\ and\ \bibinfo {author} {\bibfnamefont {S.}~\bibnamefont
  {Mankovsky}},\ }\bibfield  {title} {\enquote {\bibinfo {title} {Anisotropic
  exchange coupling in diluted magnetic semiconductors: Ab initio spin-density
  functional theory},}\ }\href@noop {} {\bibfield  {journal} {\bibinfo
  {journal} {Phys. Rev. B}\ }\textbf {\bibinfo {volume} {79}},\ \bibinfo
  {pages} {045209} (\bibinfo {year} {2009})}\BibitemShut {NoStop}%
\bibitem [{\citenamefont {Solovyev}(2021)}]{Solovyev:2021}%
  \BibitemOpen
  \bibfield  {author} {\bibinfo {author} {\bibfnamefont {I.~V.}\ \bibnamefont
  {Solovyev}},\ }\bibfield  {title} {\enquote {\bibinfo {title} {Exchange
  interactions and magnetic force theorem},}\ }\href@noop {} {\bibfield
  {journal} {\bibinfo  {journal} {Phys. Rev. B}\ }\textbf {\bibinfo {volume}
  {103}},\ \bibinfo {pages} {104428} (\bibinfo {year} {2021})}\BibitemShut
  {NoStop}%
\bibitem [{\citenamefont {Heinze}\ \emph {et~al.}(2011)\citenamefont {Heinze},
  \citenamefont {Von~Bergmann}, \citenamefont {Menzel}, \citenamefont {Brede},
  \citenamefont {Kubetzka}, \citenamefont {Wiesendanger}, \citenamefont
  {Bihlmayer},\ and\ \citenamefont {Bl{\"u}gel}}]{heinze2011spontaneous}%
  \BibitemOpen
  \bibfield  {author} {\bibinfo {author} {\bibfnamefont {S.}~\bibnamefont
  {Heinze}}, \bibinfo {author} {\bibfnamefont {K.}~\bibnamefont
  {Von~Bergmann}}, \bibinfo {author} {\bibfnamefont {M.}~\bibnamefont
  {Menzel}}, \bibinfo {author} {\bibfnamefont {J.}~\bibnamefont {Brede}},
  \bibinfo {author} {\bibfnamefont {A.}~\bibnamefont {Kubetzka}}, \bibinfo
  {author} {\bibfnamefont {R.}~\bibnamefont {Wiesendanger}}, \bibinfo {author}
  {\bibfnamefont {G.}~\bibnamefont {Bihlmayer}}, \ and\ \bibinfo {author}
  {\bibfnamefont {S.}~\bibnamefont {Bl{\"u}gel}},\ }\bibfield  {title}
  {\enquote {\bibinfo {title} {Spontaneous atomic-scale magnetic skyrmion
  lattice in two dimensions},}\ }\href@noop {} {\bibfield  {journal} {\bibinfo
  {journal} {Nature Physics}\ }\textbf {\bibinfo {volume} {7}},\ \bibinfo
  {pages} {713--718} (\bibinfo {year} {2011})}\BibitemShut {NoStop}%
\bibitem [{\citenamefont {Yao}\ \emph {et~al.}(2023{\natexlab{b}})\citenamefont
  {Yao}, \citenamefont {Gaudet}, \citenamefont {Verma}, \citenamefont {Graf},
  \citenamefont {Yang}, \citenamefont {Bahrami}, \citenamefont {Zhang},
  \citenamefont {Aczel}, \citenamefont {Subedi}, \citenamefont {Torchinsky},
  \citenamefont {Sun}, \citenamefont {Bansil}, \citenamefont {Huang},
  \citenamefont {Singh}, \citenamefont {Blaha}, \citenamefont
  {Nikoli\ifmmode~\acute{c}\else \'{c}\fi{}},\ and\ \citenamefont
  {Tafti}}]{Yao2023}%
  \BibitemOpen
  \bibfield  {author} {\bibinfo {author} {\bibfnamefont {Xiaohan}\ \bibnamefont
  {Yao}}, \bibinfo {author} {\bibfnamefont {Jonathan}\ \bibnamefont {Gaudet}},
  \bibinfo {author} {\bibfnamefont {Rahul}\ \bibnamefont {Verma}}, \bibinfo
  {author} {\bibfnamefont {David~E.}\ \bibnamefont {Graf}}, \bibinfo {author}
  {\bibfnamefont {Hung-Yu}\ \bibnamefont {Yang}}, \bibinfo {author}
  {\bibfnamefont {Faranak}\ \bibnamefont {Bahrami}}, \bibinfo {author}
  {\bibfnamefont {Ruiqi}\ \bibnamefont {Zhang}}, \bibinfo {author}
  {\bibfnamefont {Adam~A.}\ \bibnamefont {Aczel}}, \bibinfo {author}
  {\bibfnamefont {Sujan}\ \bibnamefont {Subedi}}, \bibinfo {author}
  {\bibfnamefont {Darius~H.}\ \bibnamefont {Torchinsky}}, \bibinfo {author}
  {\bibfnamefont {Jianwei}\ \bibnamefont {Sun}}, \bibinfo {author}
  {\bibfnamefont {Arun}\ \bibnamefont {Bansil}}, \bibinfo {author}
  {\bibfnamefont {Shin-Ming}\ \bibnamefont {Huang}}, \bibinfo {author}
  {\bibfnamefont {Bahadur}\ \bibnamefont {Singh}}, \bibinfo {author}
  {\bibfnamefont {Peter}\ \bibnamefont {Blaha}}, \bibinfo {author}
  {\bibfnamefont {Predrag}\ \bibnamefont {Nikoli\ifmmode~\acute{c}\else
  \'{c}\fi{}}}, \ and\ \bibinfo {author} {\bibfnamefont {Fazel}\ \bibnamefont
  {Tafti}},\ }\bibfield  {title} {\enquote {\bibinfo {title} {{Large
  Topological Hall Effect and Spiral Magnetic Order in the Weyl Semimetal
  SmAlSi}},}\ }\href@noop {} {\bibfield  {journal} {\bibinfo  {journal} {Phys.
  Rev. X}\ }\textbf {\bibinfo {volume} {13}},\ \bibinfo {pages} {011035}
  (\bibinfo {year} {2023}{\natexlab{b}})}\BibitemShut {NoStop}%
\bibitem [{\citenamefont {Yang}\ \emph
  {et~al.}(2021{\natexlab{b}})\citenamefont {Yang}, \citenamefont {Singh},
  \citenamefont {Gaudet}, \citenamefont {Lu}, \citenamefont {Huang},
  \citenamefont {Chiu}, \citenamefont {Huang}, \citenamefont {Wang},
  \citenamefont {Bahrami}, \citenamefont {Xu}, \citenamefont {Franklin},
  \citenamefont {Sochnikov}, \citenamefont {Graf}, \citenamefont {Xu},
  \citenamefont {Zhao}, \citenamefont {Hoffman}, \citenamefont {Lin},
  \citenamefont {Torchinsky}, \citenamefont {Broholm}, \citenamefont {Bansil},\
  and\ \citenamefont {Tafti}}]{Yang2021}%
  \BibitemOpen
  \bibfield  {author} {\bibinfo {author} {\bibfnamefont {Hung-Yu}\ \bibnamefont
  {Yang}}, \bibinfo {author} {\bibfnamefont {Bahadur}\ \bibnamefont {Singh}},
  \bibinfo {author} {\bibfnamefont {Jonathan}\ \bibnamefont {Gaudet}}, \bibinfo
  {author} {\bibfnamefont {Baozhu}\ \bibnamefont {Lu}}, \bibinfo {author}
  {\bibfnamefont {Cheng-Yi}\ \bibnamefont {Huang}}, \bibinfo {author}
  {\bibfnamefont {Wei-Chi}\ \bibnamefont {Chiu}}, \bibinfo {author}
  {\bibfnamefont {Shin-Ming}\ \bibnamefont {Huang}}, \bibinfo {author}
  {\bibfnamefont {Baokai}\ \bibnamefont {Wang}}, \bibinfo {author}
  {\bibfnamefont {Faranak}\ \bibnamefont {Bahrami}}, \bibinfo {author}
  {\bibfnamefont {Bochao}\ \bibnamefont {Xu}}, \bibinfo {author} {\bibfnamefont
  {Jacob}\ \bibnamefont {Franklin}}, \bibinfo {author} {\bibfnamefont {Ilya}\
  \bibnamefont {Sochnikov}}, \bibinfo {author} {\bibfnamefont {David~E.}\
  \bibnamefont {Graf}}, \bibinfo {author} {\bibfnamefont {Guangyong}\
  \bibnamefont {Xu}}, \bibinfo {author} {\bibfnamefont {Yang}\ \bibnamefont
  {Zhao}}, \bibinfo {author} {\bibfnamefont {Christina~M.}\ \bibnamefont
  {Hoffman}}, \bibinfo {author} {\bibfnamefont {Hsin}\ \bibnamefont {Lin}},
  \bibinfo {author} {\bibfnamefont {Darius~H.}\ \bibnamefont {Torchinsky}},
  \bibinfo {author} {\bibfnamefont {Collin~L.}\ \bibnamefont {Broholm}},
  \bibinfo {author} {\bibfnamefont {Arun}\ \bibnamefont {Bansil}}, \ and\
  \bibinfo {author} {\bibfnamefont {Fazel}\ \bibnamefont {Tafti}},\ }\bibfield
  {title} {\enquote {\bibinfo {title} {{Noncollinear ferromagnetic Weyl
  semimetal with anisotropic anomalous Hall effect}},}\ }\href@noop {}
  {\bibfield  {journal} {\bibinfo  {journal} {Phys. Rev. B}\ }\textbf {\bibinfo
  {volume} {103}},\ \bibinfo {pages} {115143} (\bibinfo {year}
  {2021}{\natexlab{b}})}\BibitemShut {NoStop}%
\bibitem [{\citenamefont {Wortmann}\ \emph {et~al.}(2023)\citenamefont
  {Wortmann}, \citenamefont {Michalicek}, \citenamefont {Baadji}, \citenamefont
  {Betzinger}, \citenamefont {Bihlmayer}, \citenamefont {Bröder},
  \citenamefont {Burnus}, \citenamefont {Enkovaara}, \citenamefont {Freimuth},
  \citenamefont {Friedrich}, \citenamefont {Gerhorst}, \citenamefont
  {Granberg~Cauchi}, \citenamefont {Grytsiuk}, \citenamefont {Hanke},
  \citenamefont {Hanke}, \citenamefont {Heide}, \citenamefont {Heinze},
  \citenamefont {Hilgers}, \citenamefont {Janssen}, \citenamefont
  {Klüppelberg}, \citenamefont {Kovacik}, \citenamefont {Kurz}, \citenamefont
  {Lezaic}, \citenamefont {Madsen}, \citenamefont {Mokrousov}, \citenamefont
  {Neukirchen}, \citenamefont {Redies}, \citenamefont {Rost}, \citenamefont
  {Schlipf}, \citenamefont {Schindlmayr}, \citenamefont {Winkelmann},\ and\
  \citenamefont {Blügel}}]{fleur}%
  \BibitemOpen
  \bibfield  {author} {\bibinfo {author} {\bibfnamefont {Daniel}\ \bibnamefont
  {Wortmann}}, \bibinfo {author} {\bibfnamefont {Gregor}\ \bibnamefont
  {Michalicek}}, \bibinfo {author} {\bibfnamefont {Nadjib}\ \bibnamefont
  {Baadji}}, \bibinfo {author} {\bibfnamefont {Markus}\ \bibnamefont
  {Betzinger}}, \bibinfo {author} {\bibfnamefont {Gustav}\ \bibnamefont
  {Bihlmayer}}, \bibinfo {author} {\bibfnamefont {Jens}\ \bibnamefont
  {Bröder}}, \bibinfo {author} {\bibfnamefont {Tobias}\ \bibnamefont
  {Burnus}}, \bibinfo {author} {\bibfnamefont {Jussi}\ \bibnamefont
  {Enkovaara}}, \bibinfo {author} {\bibfnamefont {Frank}\ \bibnamefont
  {Freimuth}}, \bibinfo {author} {\bibfnamefont {Christoph}\ \bibnamefont
  {Friedrich}}, \bibinfo {author} {\bibfnamefont {Christian-Roman}\
  \bibnamefont {Gerhorst}}, \bibinfo {author} {\bibfnamefont {Sabastian}\
  \bibnamefont {Granberg~Cauchi}}, \bibinfo {author} {\bibfnamefont {Uliana}\
  \bibnamefont {Grytsiuk}}, \bibinfo {author} {\bibfnamefont {Andrea}\
  \bibnamefont {Hanke}}, \bibinfo {author} {\bibfnamefont {Jan-Philipp}\
  \bibnamefont {Hanke}}, \bibinfo {author} {\bibfnamefont {Marcus}\
  \bibnamefont {Heide}}, \bibinfo {author} {\bibfnamefont {Stefan}\
  \bibnamefont {Heinze}}, \bibinfo {author} {\bibfnamefont {Robin}\
  \bibnamefont {Hilgers}}, \bibinfo {author} {\bibfnamefont {Henning}\
  \bibnamefont {Janssen}}, \bibinfo {author} {\bibfnamefont {Daniel~Aaaron}\
  \bibnamefont {Klüppelberg}}, \bibinfo {author} {\bibfnamefont {Roman}\
  \bibnamefont {Kovacik}}, \bibinfo {author} {\bibfnamefont {Philipp}\
  \bibnamefont {Kurz}}, \bibinfo {author} {\bibfnamefont {Marjana}\
  \bibnamefont {Lezaic}}, \bibinfo {author} {\bibfnamefont {Georg K.~H.}\
  \bibnamefont {Madsen}}, \bibinfo {author} {\bibfnamefont {Yuriy}\
  \bibnamefont {Mokrousov}}, \bibinfo {author} {\bibfnamefont {Alexander}\
  \bibnamefont {Neukirchen}}, \bibinfo {author} {\bibfnamefont {Matthias}\
  \bibnamefont {Redies}}, \bibinfo {author} {\bibfnamefont {Stefan}\
  \bibnamefont {Rost}}, \bibinfo {author} {\bibfnamefont {Martin}\ \bibnamefont
  {Schlipf}}, \bibinfo {author} {\bibfnamefont {Arno}\ \bibnamefont
  {Schindlmayr}}, \bibinfo {author} {\bibfnamefont {Miriam}\ \bibnamefont
  {Winkelmann}}, \ and\ \bibinfo {author} {\bibfnamefont {Stefan}\ \bibnamefont
  {Blügel}},\ }\bibfield  {title} {\enquote {\bibinfo {title} {Fleur},}\
  }\href {\doibase 10.5281/zenodo.7891361} {\  (\bibinfo {year} {2023}),\
  10.5281/zenodo.7891361}\BibitemShut {NoStop}%
\bibitem [{\citenamefont {Zhang}\ \emph {et~al.}(2023)\citenamefont {Zhang},
  \citenamefont {Gao}, \citenamefont {Gao}, \citenamefont {Lei}, \citenamefont
  {Ni}, \citenamefont {Oh}, \citenamefont {Huang}, \citenamefont {Yue},
  \citenamefont {Zonno}, \citenamefont {Gorovikov} \emph
  {et~al.}}]{zhang2023kramers}%
  \BibitemOpen
  \bibfield  {author} {\bibinfo {author} {\bibfnamefont {Yichen}\ \bibnamefont
  {Zhang}}, \bibinfo {author} {\bibfnamefont {Yuxiang}\ \bibnamefont {Gao}},
  \bibinfo {author} {\bibfnamefont {Xue-Jian}\ \bibnamefont {Gao}}, \bibinfo
  {author} {\bibfnamefont {Shiming}\ \bibnamefont {Lei}}, \bibinfo {author}
  {\bibfnamefont {Zhuoliang}\ \bibnamefont {Ni}}, \bibinfo {author}
  {\bibfnamefont {Ji~Seop}\ \bibnamefont {Oh}}, \bibinfo {author}
  {\bibfnamefont {Jianwei}\ \bibnamefont {Huang}}, \bibinfo {author}
  {\bibfnamefont {Ziqin}\ \bibnamefont {Yue}}, \bibinfo {author} {\bibfnamefont
  {Marta}\ \bibnamefont {Zonno}}, \bibinfo {author} {\bibfnamefont {Sergey}\
  \bibnamefont {Gorovikov}},  \emph {et~al.},\ }\bibfield  {title} {\enquote
  {\bibinfo {title} {Kramers nodal lines and weyl fermions in smalsi},}\
  }\href@noop {} {\bibfield  {journal} {\bibinfo  {journal} {Communications
  Physics}\ }\textbf {\bibinfo {volume} {6}},\ \bibinfo {pages} {134} (\bibinfo
  {year} {2023})}\BibitemShut {NoStop}%
\bibitem [{\citenamefont {Gubanov}\ \emph {et~al.}(2012)\citenamefont
  {Gubanov}, \citenamefont {Liechtenstein},\ and\ \citenamefont
  {Postnikov}}]{gubanov2012magnetism}%
  \BibitemOpen
  \bibfield  {author} {\bibinfo {author} {\bibfnamefont {Vladimir~A}\
  \bibnamefont {Gubanov}}, \bibinfo {author} {\bibfnamefont {Alexandr~I}\
  \bibnamefont {Liechtenstein}}, \ and\ \bibinfo {author} {\bibfnamefont
  {Andrei~V}\ \bibnamefont {Postnikov}},\ }\href@noop {} {\emph {\bibinfo
  {title} {Magnetism and the electronic structure of crystals}}},\
  Vol.~\bibinfo {volume} {98}\ (\bibinfo  {publisher} {Springer Science \&
  Business Media},\ \bibinfo {year} {2012})\BibitemShut {NoStop}%
\bibitem [{\citenamefont {Duan}\ \emph {et~al.}(2006)\citenamefont {Duan},
  \citenamefont {Sabiryanov}, \citenamefont {Mei}, \citenamefont {Dowben},
  \citenamefont {Jaswal},\ and\ \citenamefont {Tsymbal}}]{duan2006magnetic}%
  \BibitemOpen
  \bibfield  {author} {\bibinfo {author} {\bibfnamefont {Chun-Gang}\
  \bibnamefont {Duan}}, \bibinfo {author} {\bibfnamefont {Renat~F}\
  \bibnamefont {Sabiryanov}}, \bibinfo {author} {\bibfnamefont {Wai-Ning}\
  \bibnamefont {Mei}}, \bibinfo {author} {\bibfnamefont {Peter~A}\ \bibnamefont
  {Dowben}}, \bibinfo {author} {\bibfnamefont {SS}~\bibnamefont {Jaswal}}, \
  and\ \bibinfo {author} {\bibfnamefont {Evgeny~Y}\ \bibnamefont {Tsymbal}},\
  }\bibfield  {title} {\enquote {\bibinfo {title} {{Magnetic ordering in Gd
  monopnictides: Indirect exchange versus superexchange interaction}},}\
  }\href@noop {} {\bibfield  {journal} {\bibinfo  {journal} {Applied physics
  letters}\ }\textbf {\bibinfo {volume} {88}} (\bibinfo {year}
  {2006})}\BibitemShut {NoStop}%
\bibitem [{\citenamefont {Chang}\ \emph {et~al.}(2018)\citenamefont {Chang},
  \citenamefont {Singh}, \citenamefont {Xu}, \citenamefont {Bian},
  \citenamefont {Huang}, \citenamefont {Hsu}, \citenamefont {Belopolski},
  \citenamefont {Alidoust}, \citenamefont {Sanchez}, \citenamefont {Zheng},
  \citenamefont {Lu}, \citenamefont {Zhang}, \citenamefont {Bian},
  \citenamefont {Chang}, \citenamefont {Jeng}, \citenamefont {Bansil},
  \citenamefont {Hsu}, \citenamefont {Jia}, \citenamefont {Neupert},
  \citenamefont {Lin},\ and\ \citenamefont {Hasan}}]{Chang2018}%
  \BibitemOpen
  \bibfield  {author} {\bibinfo {author} {\bibfnamefont {Guoqing}\ \bibnamefont
  {Chang}}, \bibinfo {author} {\bibfnamefont {Bahadur}\ \bibnamefont {Singh}},
  \bibinfo {author} {\bibfnamefont {Su-Yang}\ \bibnamefont {Xu}}, \bibinfo
  {author} {\bibfnamefont {Guang}\ \bibnamefont {Bian}}, \bibinfo {author}
  {\bibfnamefont {Shin-Ming}\ \bibnamefont {Huang}}, \bibinfo {author}
  {\bibfnamefont {Chuang-Han}\ \bibnamefont {Hsu}}, \bibinfo {author}
  {\bibfnamefont {Ilya}\ \bibnamefont {Belopolski}}, \bibinfo {author}
  {\bibfnamefont {Nasser}\ \bibnamefont {Alidoust}}, \bibinfo {author}
  {\bibfnamefont {Daniel~S.}\ \bibnamefont {Sanchez}}, \bibinfo {author}
  {\bibfnamefont {Hao}\ \bibnamefont {Zheng}}, \bibinfo {author} {\bibfnamefont
  {Hong}\ \bibnamefont {Lu}}, \bibinfo {author} {\bibfnamefont {Xiao}\
  \bibnamefont {Zhang}}, \bibinfo {author} {\bibfnamefont {Yi}~\bibnamefont
  {Bian}}, \bibinfo {author} {\bibfnamefont {Tay-Rong}\ \bibnamefont {Chang}},
  \bibinfo {author} {\bibfnamefont {Horng-Tay}\ \bibnamefont {Jeng}}, \bibinfo
  {author} {\bibfnamefont {Arun}\ \bibnamefont {Bansil}}, \bibinfo {author}
  {\bibfnamefont {Han}\ \bibnamefont {Hsu}}, \bibinfo {author} {\bibfnamefont
  {Shuang}\ \bibnamefont {Jia}}, \bibinfo {author} {\bibfnamefont {Titus}\
  \bibnamefont {Neupert}}, \bibinfo {author} {\bibfnamefont {Hsin}\
  \bibnamefont {Lin}}, \ and\ \bibinfo {author} {\bibfnamefont {M.~Zahid}\
  \bibnamefont {Hasan}},\ }\bibfield  {title} {\enquote {\bibinfo {title}
  {Magnetic and noncentrosymmetric weyl fermion semimetals in the
  $\mathit{R}\mathrm{AlGe}$ family of compounds
  ($\mathit{R}=\mathrm{rare}\phantom{\rule{0.28em}{0ex}}\mathrm{earth}$)},}\
  }\href@noop {} {\bibfield  {journal} {\bibinfo  {journal} {Phys. Rev. B}\
  }\textbf {\bibinfo {volume} {97}},\ \bibinfo {pages} {041104} (\bibinfo
  {year} {2018})}\BibitemShut {NoStop}%
\bibitem [{\citenamefont {Gyorffy}(1972)}]{Gyorffy1972}%
  \BibitemOpen
  \bibfield  {author} {\bibinfo {author} {\bibfnamefont {B.~L.}\ \bibnamefont
  {Gyorffy}},\ }\bibfield  {title} {\enquote {\bibinfo {title}
  {Coherent-potential approximation for a nonoverlapping-muffin-tin-potential
  model of random substitutional alloys},}\ }\href@noop {} {\bibfield
  {journal} {\bibinfo  {journal} {Phys. Rev. B}\ }\textbf {\bibinfo {volume}
  {5}},\ \bibinfo {pages} {2382--2384} (\bibinfo {year} {1972})}\BibitemShut
  {NoStop}%
\bibitem [{\citenamefont {Ebert}\ \emph {et~al.}(2011)\citenamefont {Ebert},
  \citenamefont {Koedderitzsch},\ and\ \citenamefont
  {Minar}}]{ebert2011calculating}%
  \BibitemOpen
  \bibfield  {author} {\bibinfo {author} {\bibfnamefont {Hubert}\ \bibnamefont
  {Ebert}}, \bibinfo {author} {\bibfnamefont {Diemo}\ \bibnamefont
  {Koedderitzsch}}, \ and\ \bibinfo {author} {\bibfnamefont {Jan}\ \bibnamefont
  {Minar}},\ }\bibfield  {title} {\enquote {\bibinfo {title} {{Calculating
  condensed matter properties using the KKR-Green's function method—recent
  developments and applications}},}\ }\href@noop {} {\bibfield  {journal}
  {\bibinfo  {journal} {Reports on Progress in Physics}\ }\textbf {\bibinfo
  {volume} {74}},\ \bibinfo {pages} {096501} (\bibinfo {year}
  {2011})}\BibitemShut {NoStop}%
\bibitem [{\citenamefont {Patrick}\ \emph {et~al.}(2020)\citenamefont
  {Patrick}, \citenamefont {Marchant},\ and\ \citenamefont
  {Staunton}}]{patrick2020spin}%
  \BibitemOpen
  \bibfield  {author} {\bibinfo {author} {\bibfnamefont {Christopher~E}\
  \bibnamefont {Patrick}}, \bibinfo {author} {\bibfnamefont {George~A}\
  \bibnamefont {Marchant}}, \ and\ \bibinfo {author} {\bibfnamefont {Julie~B}\
  \bibnamefont {Staunton}},\ }\bibfield  {title} {\enquote {\bibinfo {title}
  {{Spin orientation and magnetostriction of Tb$_{1-x}$ Dy$_x$ Fe$_2$ from
  first principles}},}\ }\href@noop {} {\bibfield  {journal} {\bibinfo
  {journal} {Physical Review Applied}\ }\textbf {\bibinfo {volume} {14}},\
  \bibinfo {pages} {014091} (\bibinfo {year} {2020})}\BibitemShut {NoStop}%
\bibitem [{\citenamefont {M\"uller}\ \emph {et~al.}(2019)\citenamefont
  {M\"uller}, \citenamefont {Hoffmann}, \citenamefont {Di\ss{}elkamp},
  \citenamefont {Sch\"urhoff}, \citenamefont {Mavros}, \citenamefont
  {Sallermann}, \citenamefont {Kiselev}, \citenamefont {J\'onsson},\ and\
  \citenamefont {Bl\"ugel}}]{Gideon:2019}%
  \BibitemOpen
  \bibfield  {author} {\bibinfo {author} {\bibfnamefont {Gideon~P.}\
  \bibnamefont {M\"uller}}, \bibinfo {author} {\bibfnamefont {Markus}\
  \bibnamefont {Hoffmann}}, \bibinfo {author} {\bibfnamefont {Constantin}\
  \bibnamefont {Di\ss{}elkamp}}, \bibinfo {author} {\bibfnamefont {Daniel}\
  \bibnamefont {Sch\"urhoff}}, \bibinfo {author} {\bibfnamefont {Stefanos}\
  \bibnamefont {Mavros}}, \bibinfo {author} {\bibfnamefont {Moritz}\
  \bibnamefont {Sallermann}}, \bibinfo {author} {\bibfnamefont {Nikolai~S.}\
  \bibnamefont {Kiselev}}, \bibinfo {author} {\bibfnamefont {Hannes}\
  \bibnamefont {J\'onsson}}, \ and\ \bibinfo {author} {\bibfnamefont {Stefan}\
  \bibnamefont {Bl\"ugel}},\ }\bibfield  {title} {\enquote {\bibinfo {title}
  {Spirit: Multifunctional framework for atomistic spin simulations.}}\
  }\href@noop {} {\bibfield  {journal} {\bibinfo  {journal} {Phys. Rev. B}\
  }\textbf {\bibinfo {volume} {99}},\ \bibinfo {pages} {224414} (\bibinfo
  {year} {2019})}\BibitemShut {NoStop}%
\bibitem [{\citenamefont {Bouaziz}\ \emph {et~al.}(2021)\citenamefont
  {Bouaziz}, \citenamefont {Ishida}, \citenamefont {Lounis},\ and\
  \citenamefont {Bl\"ugel}}]{Bouaziz2021}%
  \BibitemOpen
  \bibfield  {author} {\bibinfo {author} {\bibfnamefont {J.}~\bibnamefont
  {Bouaziz}}, \bibinfo {author} {\bibfnamefont {H.}~\bibnamefont {Ishida}},
  \bibinfo {author} {\bibfnamefont {S.}~\bibnamefont {Lounis}}, \ and\ \bibinfo
  {author} {\bibfnamefont {S.}~\bibnamefont {Bl\"ugel}},\ }\bibfield  {title}
  {\enquote {\bibinfo {title} {Transverse transport in two-dimensional
  relativistic systems with nontrivial spin textures},}\ }\href@noop {}
  {\bibfield  {journal} {\bibinfo  {journal} {Phys. Rev. Lett.}\ }\textbf
  {\bibinfo {volume} {126}},\ \bibinfo {pages} {147203} (\bibinfo {year}
  {2021})}\BibitemShut {NoStop}%
\end{thebibliography}%

\end{document}